\documentclass[11pt, oneside]{article} 
\pdfoutput=1

\usepackage{jheppub}

\usepackage{amsmath,amssymb,amsfonts,amsthm}
\usepackage{color}
\usepackage{booktabs}
\definecolor{darkblue}{rgb}{0.1,0.1,.7}
\usepackage[]{amsmath}
\usepackage[]{graphicx}
\usepackage[]{latexsym}
\usepackage{amscd}
\usepackage[all,cmtip]{xy}
\usepackage{mathrsfs}
\usepackage{bbold}
\usepackage{subfigure}
\usepackage[margin=10pt,font=small,labelfont=bf]{caption}
\usepackage{simplewick}
\usepackage{changepage}
\usepackage[]{algorithm2e}
\usepackage{booktabs,multirow}
\newtheorem*{definition*}{Definition}
\theoremstyle{remark}

\usepackage{nicematrix}
\NiceMatrixOptions{cell-space-limits = 4pt}
\makeatletter
\def\@fpheader{\ }
\makeatother
\usepackage{letltxmacro}
\makeatletter
\let\oldr@@t\r@@t
\def\r@@t#1#2{%
\setbox0=\hbox{$\oldr@@t#1{#2\,}$}\dimen0=\ht0
\advance\dimen0-0.2\ht0
\setbox2=\hbox{\vrule height\ht0 depth -\dimen0}%
{\box0\lower0.4pt\box2}}
\LetLtxMacro{\oldsqrt}{\sqrt}
\renewcommand*{\sqrt}[2][\ ]{\oldsqrt[#1]{#2}}
\makeatother

\title{Towards 3D CFT Cartography with the Stress Tensor Bootstrap}
\author{Rajeev S. Erramilli$^N$, Matthew S. Mitchell$^S$}
\affiliation{$^N$Institut des Hautes \'Etudes Scientifiques, 91440 Bures-sur-Yvette, France}
\affiliation{$^S$Department of Physics, University of Pisa and INFN, Largo Pontecorvo 3, I-56127 Pisa, Italy}

\abstract{
  We present new numerical results on the space of local, unitary, parity-preserving conformal field theories (CFTs) in three dimensions from the stress tensor bootstrap. In bounds maximizing certain OPE coefficients, we find a plethora of sharp features, such as kinks and ridges, as a function of scaling dimensions. We show that some of these features correspond to known theories, but there are many others that are equally strong but do not match known CFTs. We argue that these features are robust to raising numerical order and could then correspond to numerous as yet unknown CFTs. We conclude in proposing a program of ``CFT cartography'': the systematic exploration of the landscape of CFTs without individual theory targets in mind.}

\begin{document}

\maketitle
 
\newpage
\section{Introduction}

The conformal bootstrap program seeks to sharpen our understanding of conformal field theories (CFTs) through constraints on correlation functions deriving from the associativity of operator product expansions (OPEs) and unitarity. Over the nearly two decades of its modern revival, it has proven to be tremendously productive. The bootstrap has been successful in generating world-record bounds on the observables of physically important CFTs, such as the critical exponents of the 3D Ising model  \cite{Kos:2014bka,Kos:2016ysd,Chang:2024whx}, as well as in deepening our understanding of CFTs and QFTs writ large---see \cite{Poland:2018epd,Rychkov:2023wsd} for recent reviews. 

Among this wealth of results are \emph{universal} constraints on 3D CFTs derived from the four-point correlation function of the stress-energy tensor \(\langle TTTT\rangle\) in 2017 \cite{Dymarsky:2017yzx}. Unlike other numerical bootstrap results, which target specific CFTs, the stress tensor bootstrap's results apply to every local, unitary CFT in three dimensions.\footnote{The results presented in \cite{Dymarsky:2017yzx}, as well as those in this paper, are specific to parity-preserving CFTs. However, the stress tensor bootstrap could straightforwardly be extended to parity-violating theories.} Every such CFT is guaranteed to have a stress tensor and thus must have a consistent stress tensor four-point function. However, due to the computational cost, these constraints were not pursued further.\footnote{We should note that a related kind of constraint derived from four-point functions of global symmetry currents have been studied much more thoroughly \cite{Dymarsky:2017xzb,Reehorst:2019pzi,He:2023ewx}, even in four dimensions \cite{Karateev:2025sjw}. Any CFT with the associated global symmetry is guaranteed to have such a current. Stress tensor calculations are quite significantly more challenging, hence the disparity of interest.}

In the intervening years, the technology of the bootstrap has been developed to a significant degree. Both new analytic results and computational tools have made attainable what was previously impractical \cite{Erramilli:2019njx,Erramilli:2020rlr,Chester:2019ifh,Landry:2019qug,Chester:2020iyt,Erramilli:2022kgp,Rychkov:2023wsd}. It was with this opportunity that we (as part of a larger collaboration) were able to revisit the stress tensor bootstrap with the goal of targeting the Ising CFT \cite{Chang:2024whx}. Along the way to the Ising results in that work, we independently computed further bounds for the pure stress tensor bootstrap, some of which were reported in that previous work. In the present work, we present more of our results and expand the scope again to mapping the space of all local, unitary CFTs in 3D. Our results build upon prior literature and uncover remarkable new features on the space of CFT bounds. These features appear to be numerically stable, but we are unable to identify them with any known CFTs. As such, our results have a cartographic quality to them: to abuse the extended geographic metaphors endemic to the conformal bootstrap, we claim to discover new mountains, ridges, and valleys on the vast CFT continent.

In section \ref{sec:stress-review}, we briefly review the relevant technical details of the stress tensor bootstrap and prior results. We proceed into the bulk of our results in section \ref{sec:ttp}, focusing on bounds derived from maximizing the \(\lambda_{TT+}\) OPE coefficient. Afterwards, in section \ref{sec:other} we present miscellaneous other results derived from further assumptions or extremizations. We conclude and argue for a further program of ``CFT cartography'' in section \ref{sec:outlook}.

\section{Stress tensor bootstrap review and bounds on scalars}
\label{sec:stress-review}
\subsection{Bootstrap setup, parameters, tensor structures}
As noted in the introduction, we will be studying the bootstrap bounds derived from the consistency of four-point correlation functions of the stress-energy tensor \(\langle TTTT\rangle\). We assume that they are invariant under conformal, parity, and permutation symmetries as well as Ward identities for the stress tensor's conservation. This system has been studied previously in \cite{Dymarsky:2017yzx,Chang:2024whx}, so we refer the interested reader to the relevant works for further technical details (as well as appendix \ref{sec:params}, which lists our numerical parameters). Here we will only review prerequisite knowledge to understand our main results. Notably, we will elide descriptions of four-point tensor structures and the crossing equations themselves.

In what follows, we will be imposing gap assumptions on various OPE channels and extremizing OPE coefficients. Which channels and which (or rather, how many) coefficients appear are determined by the invariant three-point tensor structures of our symmetries and conservation conditions. These are summarized in this table:

\begin{table}[h!]
  \centering
  \begin{tabular}{c | l|c}
    Parity & Spin & \# \\
    \hline
    $+$ & \(\ell=0\) & 1\\
    $+$ & \(\ell=2\) & 1\\
    $+$ & \(\ell\geq 4,~\ell\) even&2\\
    $-$ & \(\ell=0\) & 1\\
    $-$ & \(\ell=2\) & 1\\
    $-$ & \(\ell\geq 4\) & 1
  \end{tabular}
  \caption{The parity, spin, and coefficient multiplicity of exchanged operators in the \(T\times T\) OPE, with the exception of $T$ itself.}
\end{table}

Not included in the table is the special case of the \(\langle TTT\rangle\) three-point structures i.e.\ the OPE exchange of the stress tensor itself which is of particular importance. There are two independent invariant tensors which can be seen in the following convenient parametrization of the three point function:
\begin{equation}
  \langle TTT\rangle = \lambda_B \langle TTT\rangle_B + \lambda_F \langle TTT\rangle_F
\end{equation}
where \(\langle TTT\rangle_B\) and \(\langle TTT\rangle_F\) are the three point functions of the free scalar and Majorana fermion theories, respectively. In order to define the coefficients, we must first specify that we are working with a unit-normalized stress tensor, following the conventions of \cite{Chang:2024whx}. Notably, the central charge \(c_T\) appears not as a normalization of the two-point function \(\langle TT\rangle\) but as a factor in the three-point functions \(\langle TT\mathcal{O}\rangle\). The \(\langle TTT\rangle\) three-point function coefficients satisfy
\begin{equation}
  \lambda_B + \lambda_F = \sqrt{\frac{c_B}{c_T}}
\end{equation}
where \(c_B\) is the free scalar theory's central charge. The average null energy condition (ANEC) implies the so-called conformal collider bounds \(\lambda_B\geq 0\) and \(\lambda_F\geq 0\); saturation in turn is only satisfied for the free theories themselves \cite{Maldacena:2011jn,Meltzer:2018tnm}. As the coefficients \(\lambda_B,\lambda_F\) are not both independent of the central charge \(c_T\), it is also convenient to define the parameter
\begin{equation}
  \theta \equiv \tan^{-1}\frac{\lambda_F}{\lambda_B}
\end{equation}
where \(\theta=\frac{\pi}{2}\iff \lambda_B=0\) and \(\theta=0\iff \lambda_F=0\); the conformal collider bounds are thus \(\theta\in[0,\frac{\pi}{2}]\). In the numerical bootstrap computations, we can additionally assume a specific angle \(\theta\) in the same manner as what is assumed with the OPE scan and cutting surface algorithm searches of other bootstrap studies \cite{Kos:2016ysd,Chester:2019ifh,Rychkov:2023wsd}, but unless stated otherwise we will not.

Note that with these constraints, we cannot access any spin-1 operators due to selection rules nor any parity-even spin-odd operators. Relatedly, as the stress tensor is a global symmetry singlet, we cannot access any nontrivially charged operators. Nonetheless, the bounds we derive are necessary for any theory to satisfy.

\subsection{A review of universal scalar bounds and known CFTs}

\begin{figure}
  \centering
  \includegraphics[width=0.95\textwidth]{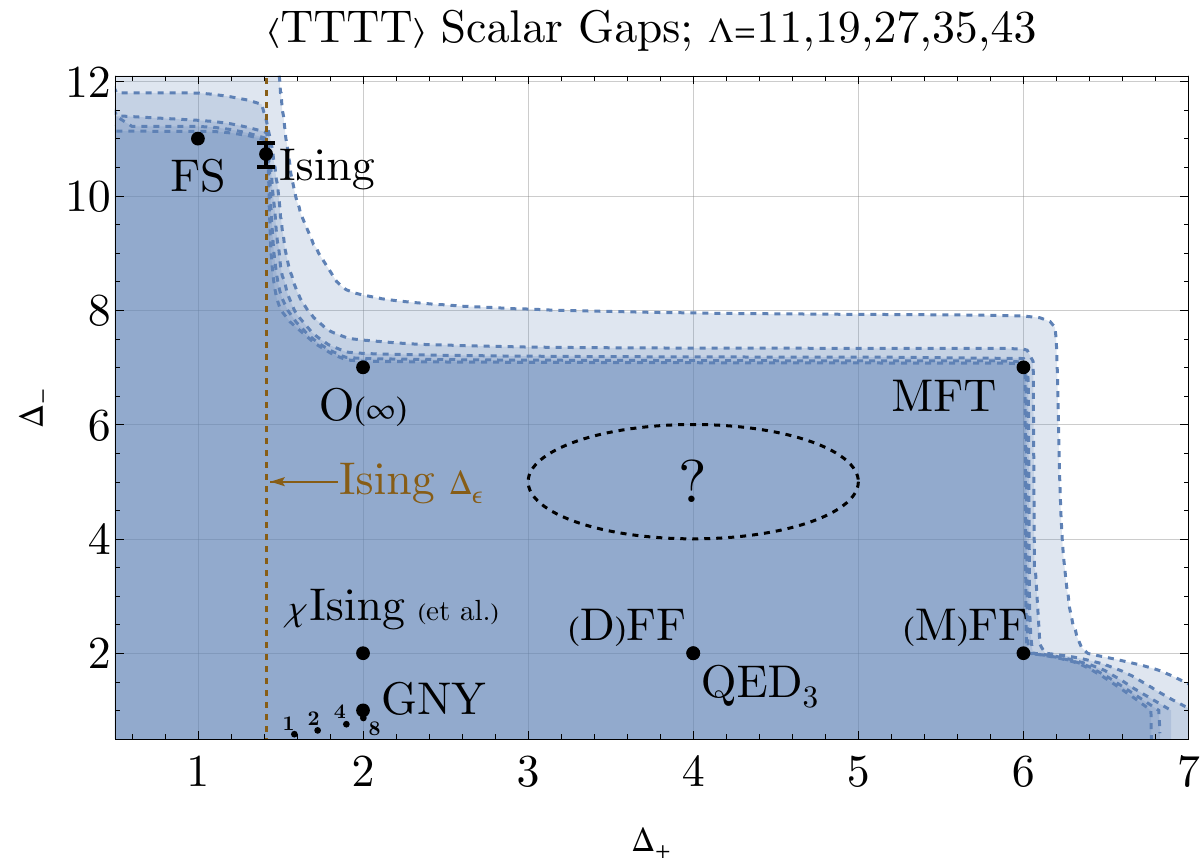}
  \caption{Allowed values of the parity-even and parity-odd singlet scalar gaps for all 3d CFTs, computed by bootstrapping on the $\langle TTTT \rangle$ correlator. 
  See main text for the labeling of points, and table \ref{tab:params1} for our solver parameters.}
  \label{fig:tttt-allowed}
\end{figure}

The system laid out in the prior section has already been studied \cite{Dymarsky:2017yzx,Chang:2024whx} to an extent in constraining the spectral gaps and the \(\langle TTT\rangle\) coefficients. The headline results were exclusion plots with respect to the gaps in parity-even and -odd scalar sectors, reproduced in figure \ref{fig:tttt-allowed}. Every local unitary 3D CFT fits somewhere into the finite allowed region of this plot. We note some special features along the boundary that correspond to known CFTs:
\begin{itemize}
  \item In the upper right corner, at (6,7), is the so-called \textbf{stress tensor mean field theory (MFT)}. This trivial and free solution to the constraints corresponds to a decoupling of the stress tensor, i.e.\ the vanishing of OPE coefficients. Equivalently, we can see this as the infinite central charge limit, which is manifest in our normalization convention.
  \item In the far upper left is the theory describing a \textbf{free real scalar} at (1,11), denoted by \textbf{FS}.\footnote{The scalar itself has $\Delta_\phi = 1/2$, but because the theory has a $\mathbb Z_2$ symmetry, the lightest singlet is $\phi^2$ with $\Delta_+ = 1$. The lightest parity-odd operator, meanwhile, is $\epsilon^{\mu\nu\rho} \phi (\partial_\alpha \partial_{\beta_1} \partial_{\beta_2} \partial_\mu \phi) (\partial^\alpha \partial_\nu \phi) (\partial^{\beta_1} \partial^{\beta_2} \partial_\rho \phi)$ (plus total derivative terms), which has $\Delta_- = 11$.}
  \item In the far lower right is the \textbf{(Majorana) free fermion theory}, denoted by \textbf{(M)FF}. While both the free Majorana and Dirac theories have a parity-odd bilinear $\psi \psi$ with $\Delta_- = 1$, the operator $(\psi\psi)^2$ vanishes in the Majorana case due to Fierz identities, so the lightest parity-even operator is \(T^2\) with $\Delta_+ = 6$.
  \item Next to the scalar free theory is a sharp kink in the bound which corresponds almost exactly to the \textbf{3D Ising CFT}, which has scaling dimensions $\Delta_+ = 1.41262528(\mathbf{29})$ and $\Delta_- \lesssim 10.93$ \cite{Chang:2024whx}.\footnote{The bounds for $\Delta_+$ are rigorous, as is the upper bound for $\Delta_-$. By making increasingly speculative assumptions about the value of $\Delta_-'$, we can obtain lower bounds on $\Delta_-$ ranging from 10.54 to 10.88.}
  \item Following the bound down from the Ising CFT, we have a curving trajectory that ultimately meets the large-\(N\) limit of \textbf{the \(O(N)\) archipelago} at $(2,7)$, denoted by \textbf{O($\infty$)}.
\end{itemize}

It's somewhat surprising that these known CFTs apparently correspond to nontrivial features in universal bounds from the consistency of the stress-tensor 4pt function. That is, it's surprising to see such a specific correspondence without more precise and theory-specific assumptions. The success of the 3D Ising bootstrap, for example, has depended on imposing targeted gap assumptions in both \(\mathbb{Z}_2\) even and odd scalar channels, and indeed this is the only way to get the iconic bootstrap islands (as far as we know). Of course, the 3D Ising bootstrap began with the discovery of kinks in bounds of \(\mathbb{Z}_2\) even and odd scalar CFT data for an otherwise relatively generic correlator of scalars. We can therefore see the stress tensor bootstrap's 3D Ising kink as similar in spirit but instead relating to \emph{parity} even and odd scalar CFT data. Still, that we need not specify the existence of \emph{any} \(\mathbb{Z}_2\) odd scalar remains remarkable. It is also noteworthy that it connects to the \(O(N)\) archipelago, which lives along the boundary of these bounds.

With that said, these correspondences along the boundary are clearly limited. What about the rest of the CFTs, which necessarily must be in the interior (if not deep interior) of this bound? We will list several known examples, which are also marked in the figure \ref{fig:tttt-allowed}:
\begin{itemize}
\item The \textbf{free Dirac fermion} is located at $(4,2)$, denoted by \textbf{(D)FF}. (Here, $(\psi^\dagger \psi)^2$ is nonvanishing, unlike in the Majorana case.)
\item The large-$N_f$ limit of \textbf{conformal QED$_3$} is located at the same point, with the $1/N_f$ corrections extending to the lower right \cite{DiPietro:2015taa,Albayrak:2021xtd,Chester:2016wrc}.
\item \textbf{Gross--Neveu--Yukawa CFTs} are in the lower left, with the large-$N$ point at (2,1), denoted by \textbf{GNY}, and the finite-$N$ archipelago extending further down and to the left. The values of $\Delta_+$ and $\Delta_-$ have been precisely determined for several $N$ via the conformal bootstrap \cite{Atanasov:2022bpi,Erramilli:2022kgp,Mitchell:2024hix}.
\item A close cousin of the GNY model is the \textbf{chiral Ising model} \cite{Rosenstein:1993zf,Gracey:1990wi,vojta2000quantum,vojta2003quantum,Moon:2012rx,Herbut:2014lfa,fei2016yukawa,Mihaila:2017ble,Zerf:2017zqi}, in which the order parameter $\phi$ is $\mathbb Z_2$-odd and parity-even---see section 2.4 and appendix C of \cite{Erramilli:2022kgp} for a more detailed explanation.  The lightest parity-odd singlet scalar is therefore the fermion bilinear $\psi^\dagger_a \psi^a$, so the large-$N$ point is at $(2,2)$, denoted by \textbf{$\chi$Ising}. There are three more associated families with the same large-$N$ point: the critical \textbf{chiral XY}, \textbf{chiral Heisenberg}, and \textbf{orthogonal Heisenberg} models, where $\phi$ is a vector under an $O(2)$ or $O(3)$ flavor symmetry \cite{Mitchell:2025oao}.
\end{itemize}

Is it possible that the stress tensor bootstrap ``knows'' about these theories as well? As the sharp features appear at boundaries, we should therefore look for other parameters to extremize beyond just the scalar gaps. We should look for other boundaries to map out, in other words. Separately, the deep interior of the bound in figure \ref{fig:tttt-allowed} leaves a large chunk of parameter space (denoted by \textbf{?}) without any known CFTs. Does the stress tensor bootstrap have anything to say here, as well?

\section{Maximizing the leading parity-even singlet scalar OPE coefficient}
\label{sec:ttp}
The simplest CFT data extremize with the numerical bootstrap are OPE coefficients as they can directly be made the objective of our numerical optimization. In this section, we will focus entirely on maximizing a single OPE coefficient, that of the leading parity-even singlet scalar \(\lambda_{TT+}\). In plain language, we ask ``What is the largest value of \(\lambda_{TT+}\) that allows for a consistent OPE?" We will compute plots assuming various spectral gaps as a function of further gap assumptions. \textit{A priori}, this sort of exploration might seem unproductive for finding sharp results, but will find a rich structure in the bounds nonetheless.

\subsection{In the neighborhood of 3D Ising}

\begin{figure}
  \centering
  \includegraphics[width=\textwidth]{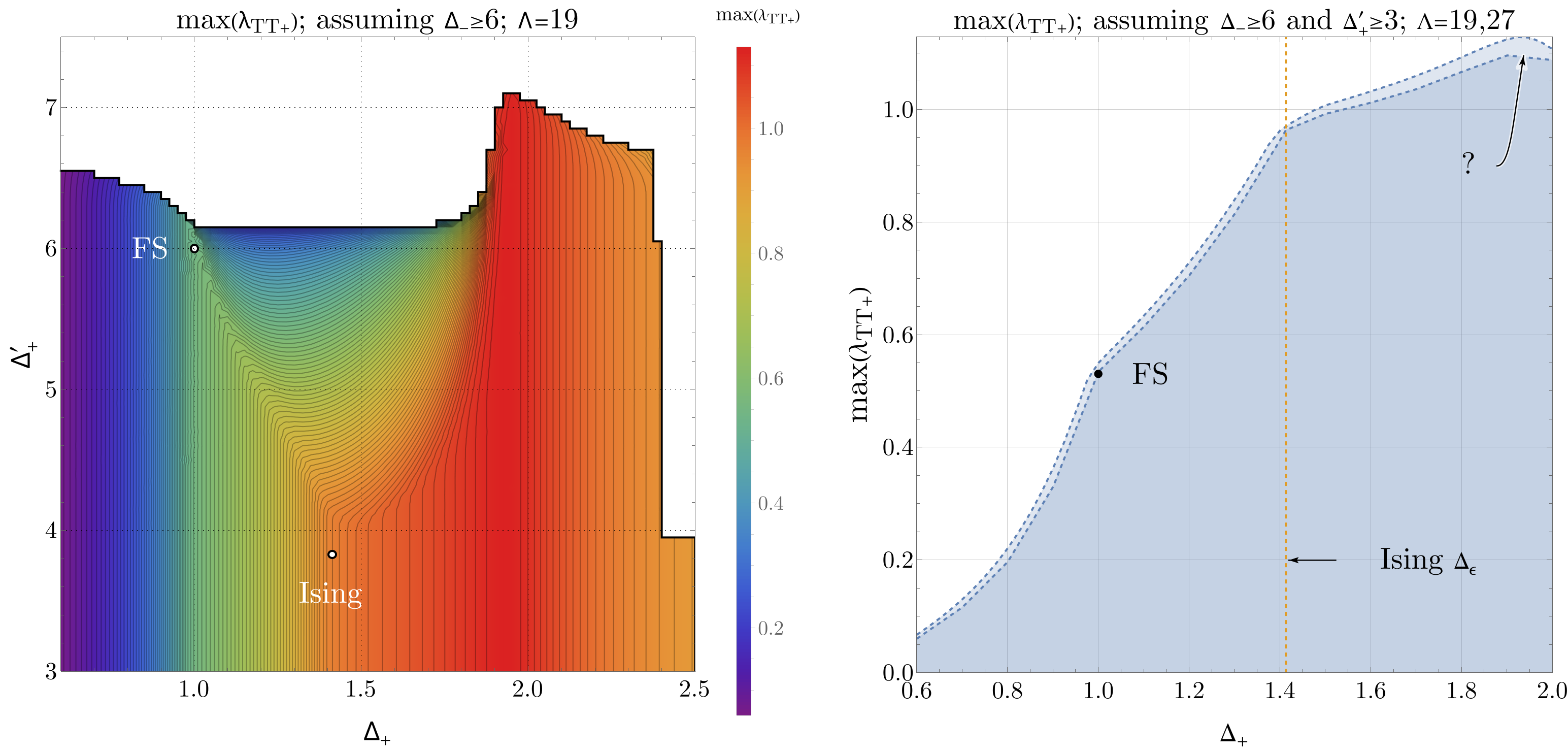}
  \caption{Results from maximizing the OPE coefficient $\lambda_{TT+}$, which corresponds to the lightest parity even scalar, reproduced from \cite{Chang:2024whx}. In the left-hand plot, we vary the gap $\Delta'_+$ to the next parity even scalar, while in the right-hand plot, we show a slice for $\Delta'_+ \ge 3$ to more clearly illustrate the kinks---one corresponding the free scalar, another corresponding to the Ising model, and a third, which remains unidentified.}
  \label{fig:max-lttp-side-by-side}
\end{figure}

To start, we will ground ourselves by looking at bounds in the neighborhood of the 3D Ising CFT. In figure \ref{fig:max-lttp-side-by-side} we reproduce figure 6 from \cite{Chang:2024whx} showing the upper bound of the leading parity-even scalar OPE coefficient as a function of the leading parity-even scalar's dimension (assuming gaps to the subleading even scalar and leading odd scalar). As was pointed out in that work, the bound shows three distinct kinks. The first corresponds exactly with the free scalar theory, the second was found to correspond exactly with the 3D Ising CFT, and the third remains somewhat mysterious. As we vary the gap to the subleading even scalar, we see that these kinks are actually part of a larger structure in the bounds; the mysterious ``?'' feature appears to be the end of a large, curving ridge which begins with the Ising kink. Even leaving that unknown feature aside, it's remarkable that at least two known theories can be located precisely with just constraints derived from the stress tensor. That is, even without knowing what the Ising CFT was or what its various lattice or Lagrangian descriptions and definitions were, we could fairly accurately (if speculatively) determine several of the CFT data! Is this unique to the Ising CFT, or are there other CFTs we could find in this manner?

\subsection{Zooming out to all 3D (local, unitary) CFTs}

\begin{figure}
  \centering
  \includegraphics[width=0.7\textwidth]{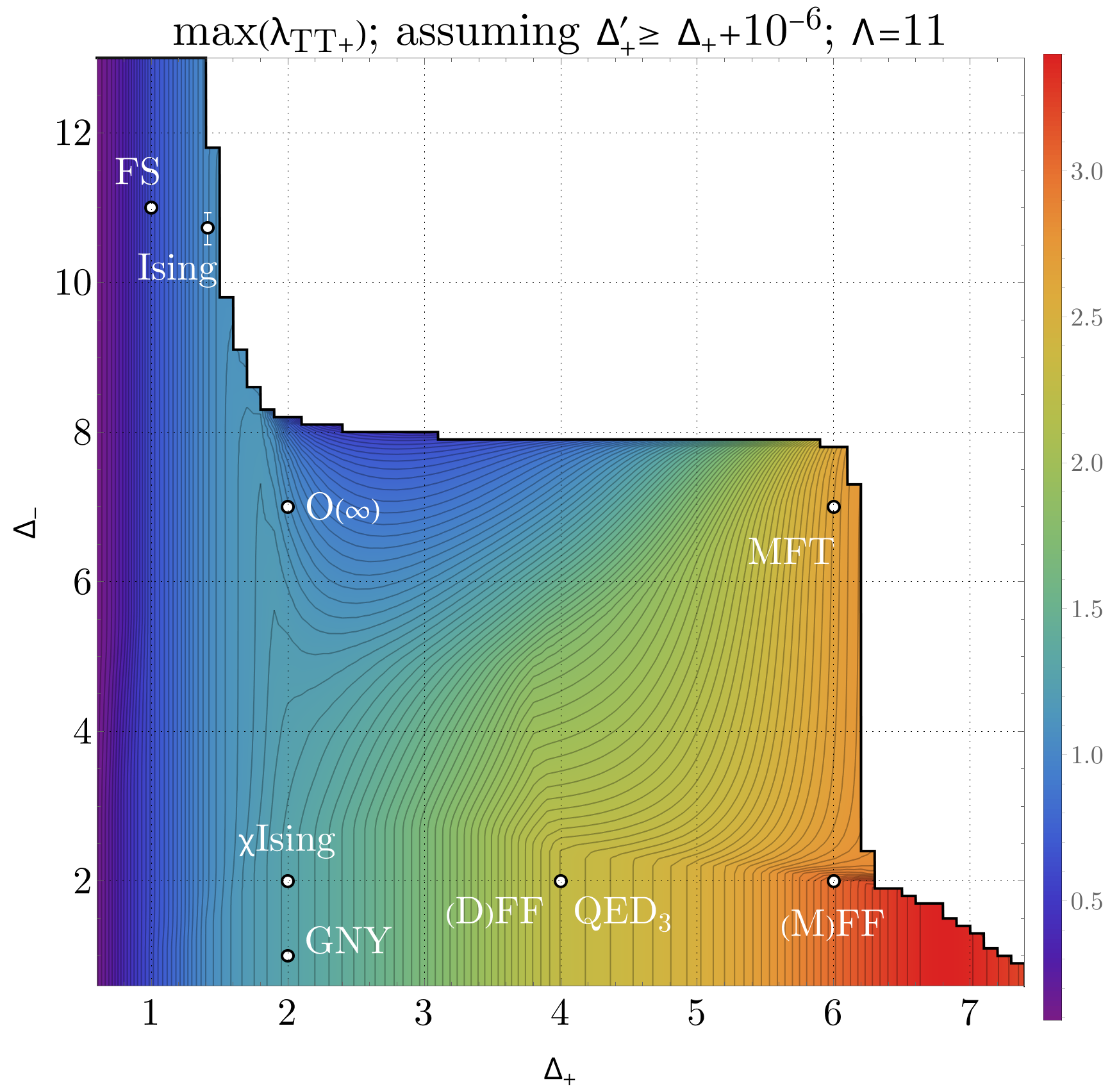}
  \caption{Contour plot, maximizing the coefficient $\lambda_{TT+}$ across the entire allowed region. Several kinks are visible, running between identified theories (see figure \ref{fig:max-lttp-dl}).}
  \label{fig:max-lttp-contour}
\end{figure}

\begin{figure}
  \centering
  \includegraphics[width=0.645\textwidth]{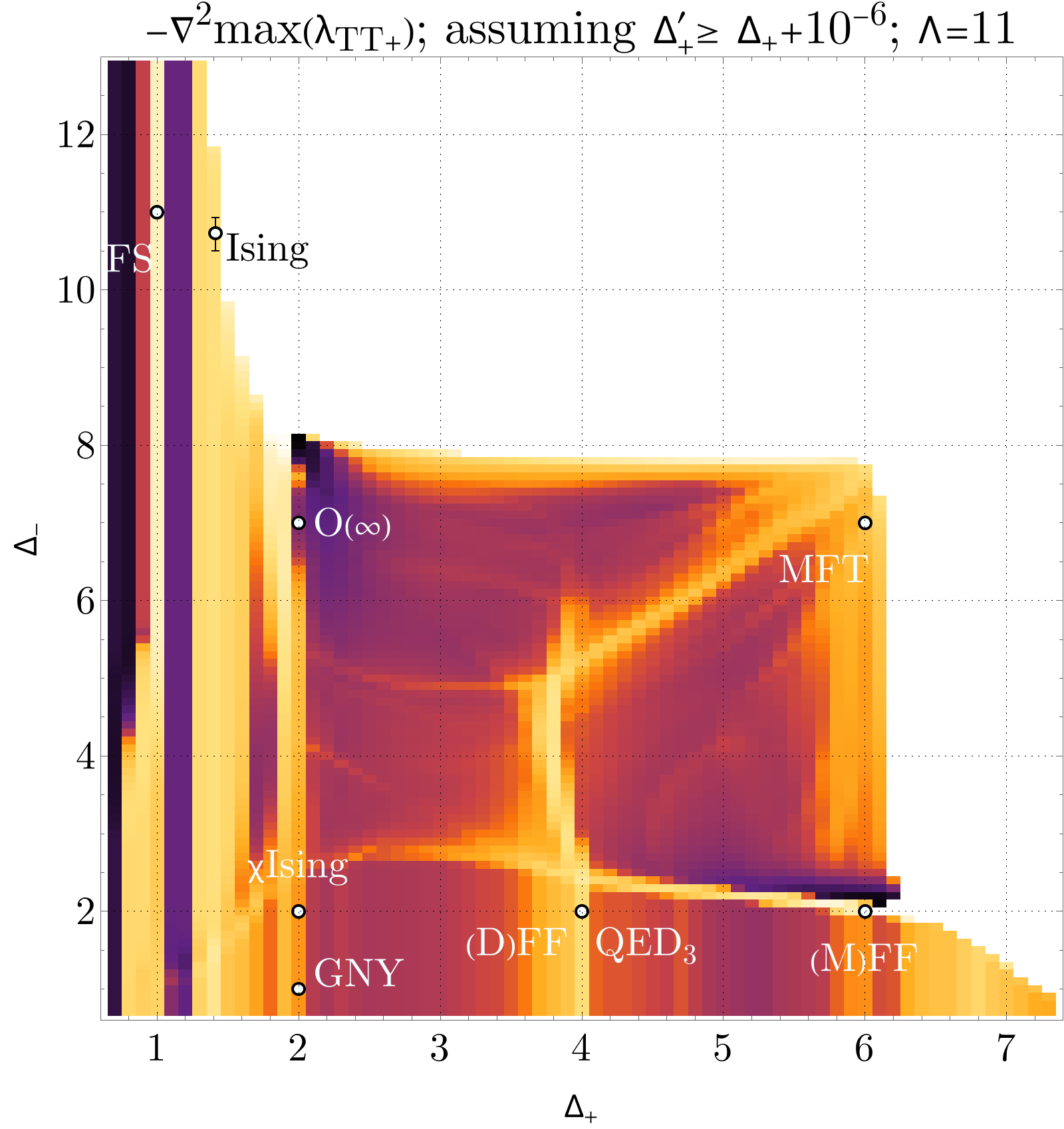}
  \caption{Discrete Laplacian of the maximal $\lambda_{TT+}$, corresponding to the curvature of the surface in figure \ref{fig:max-lttp-contour}. We see a kink running from the free Majorana fermion, to the free Dirac fermion/QED$_3$, to the chiral Ising model---we call this the ``Majorana ridge.'' Another, which we call the ``Dirac ridge,'' runs from the free Dirac fermion to a point at $(4,5)$, while a third, which we call the ``MFT ridge,'' runs through this point, up to the tensor mean field theory at $(6,7)$. (While it doesn't quite pass through this point, this is due to the slow convergence of the \(\Delta_-\) upper bound with numerical order.)}
  \label{fig:max-lttp-dl}
\end{figure}

Up until now, we have presented existing results. Taking inspiration from the kinks obtained in figure \ref{fig:max-lttp-side-by-side}, we set out to see if our OPE coefficient bound has other nontrivial features, particularly in broader parts of parameter space with more general assumptions. In figure \ref{fig:max-lttp-contour} we present the leading even scalar OPE coefficient bound at \(n_\text{max}=6\) as a function of that leading even scalar's dimension ($\Delta_+$) and the gap to the leading odd scalar ($\Delta_-$). Note the asymmetry in the meaning of the parameters. To be explicit, we are requiring the existence of an even scalar of dimension \(\Delta_+\) but disallowing the existence of \emph{any} odd scalars of dimension \(<\Delta_-\). It could be that the leading parity odd scalar has some dimension higher than \(\Delta_-\). A consequence is that the bounds are monotonically decreasing as a function of \(\Delta_-\) but are not monotonic in any sense as a function of \(\Delta_+\).\footnote{More directly, since the \(\langle TT+\rangle\) coefficient we are maximizing is itself defined in terms of \(\Delta_+\), it should be clear that monotonicity would not apply here.} Note that we no longer make any further assumptions about the scaling dimension of the subleading even scalar, just that it is, well, subleading. Figure \ref{fig:max-lttp-contour} shows hints that there is some structure to this bound in the form of some ridges and kinks in the contours.

To better see those ridges and kinks, in figure \ref{fig:max-lttp-dl} we present the discrete Laplacian of the same data as a function of the two parameters \((\Delta_+,\Delta_-)\). Bright areas (with large values of negative Laplacian) correspond to sharply concave-down sections of the OPE coefficient bound---i.e.\ kinks. The discrete Laplacian plot reveals numerous sharp and dramatic features across the parameter space that were otherwise invisible in the original contour plot, as well as lesser features that the resolution of this plot does not allow us to see. Curiously, many of these ridges indeed line up with our expectations of certain known CFTs.

Starting our tour in the far upper left, we first see a sharp vertical line up to the free scalar theory (\textbf{FS}). This is this plot's avatar of the free scalar kink found in figure \ref{fig:max-lttp-side-by-side}. Below an odd scalar gap of 6, we see that the kink is no longer as isolated and grows broader; it's unclear as to what this may correspond to. Slightly to the right of the free scalar line is what appears to be the 3D Ising kink, but beyond that, this plot lacks the resolution required to better interpret what is going on in that region.

Turning our attention to the the lower right there is a strong mostly horizontal kink or ridge starting from the Majorana free fermion theory (\textbf{(M)FF}) that is striking enough to be visible even without the discrete Laplacian. Nearby, there is a vertical ridge in the vicinity of the Dirac free fermion theory. The straightness of this ridge is a result of the two theories being the same bootstrap solution for different values of the odd scalar gap. In other words, the leading parity odd scalar is not yet saturating its gap assumption. Where this straight vertical ridge meets the ridge from the Majorana free fermion, we have an intersection that dramatically changes the nature of each. The ``Majorana'' ridge continues to the left curving gently to a maximum parity odd scalar gap of roughly 2.8 before swinging downward into a region of insufficient resolution. The ``Dirac'' ridge ceases to be straight and skews to the left as it swings upward; at an odd scalar gap of roughly 5 a blurry and amorphous diagonal line splits off to head toward the stress tensor MFT solution.

\subsection{Mysterious features near the complex free fermion}

\begin{figure}
  \centering
  \includegraphics[width=0.6\textwidth]{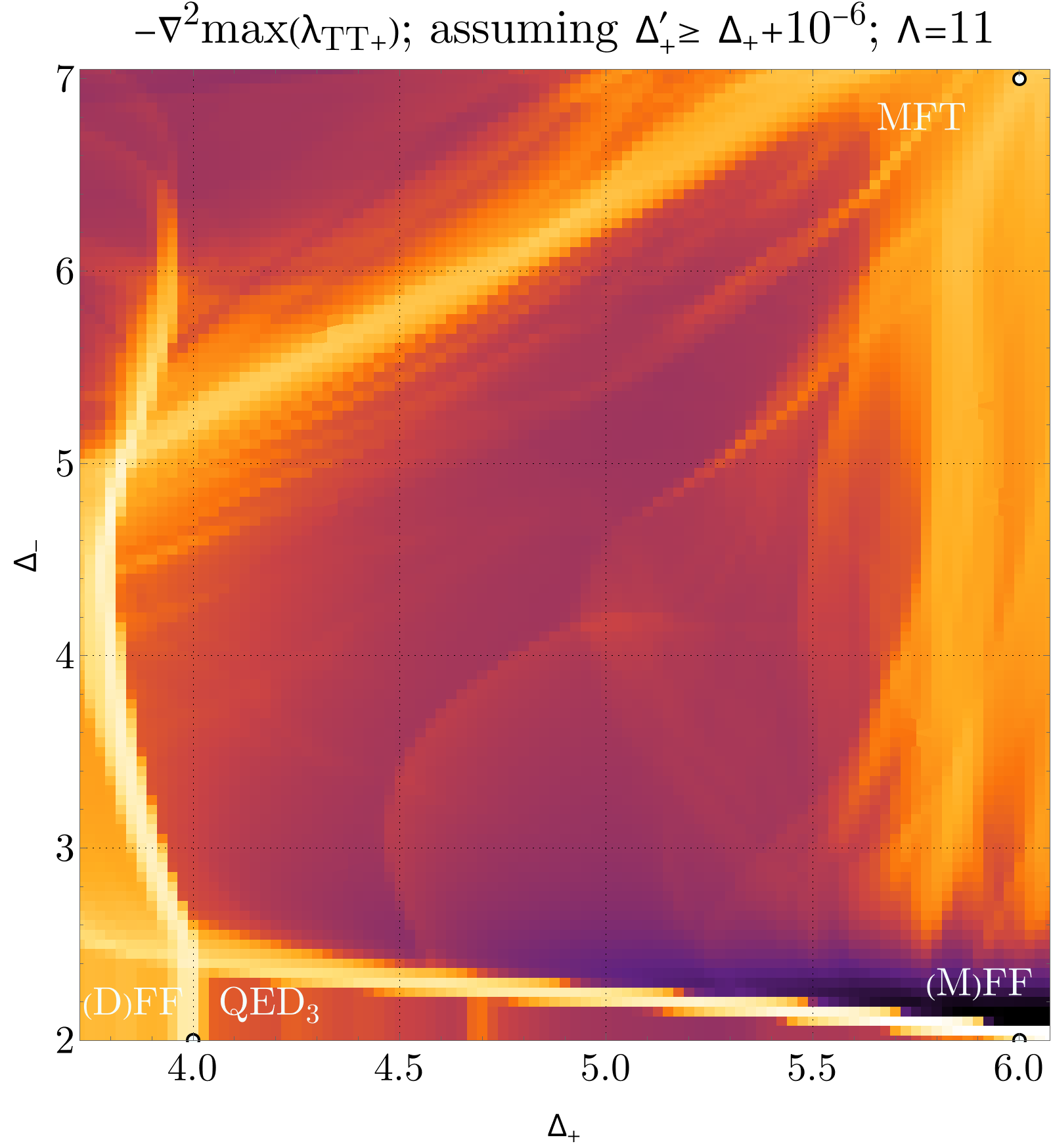}
  \caption{Higher-resolution Laplacian plot, zoomed in on the Majorana, Dirac, and MFT ridges.}
  \label{fig:max-lttp-dl-zoom1}
\end{figure}

\begin{figure}
  \centering
  \includegraphics[width=0.6\textwidth]{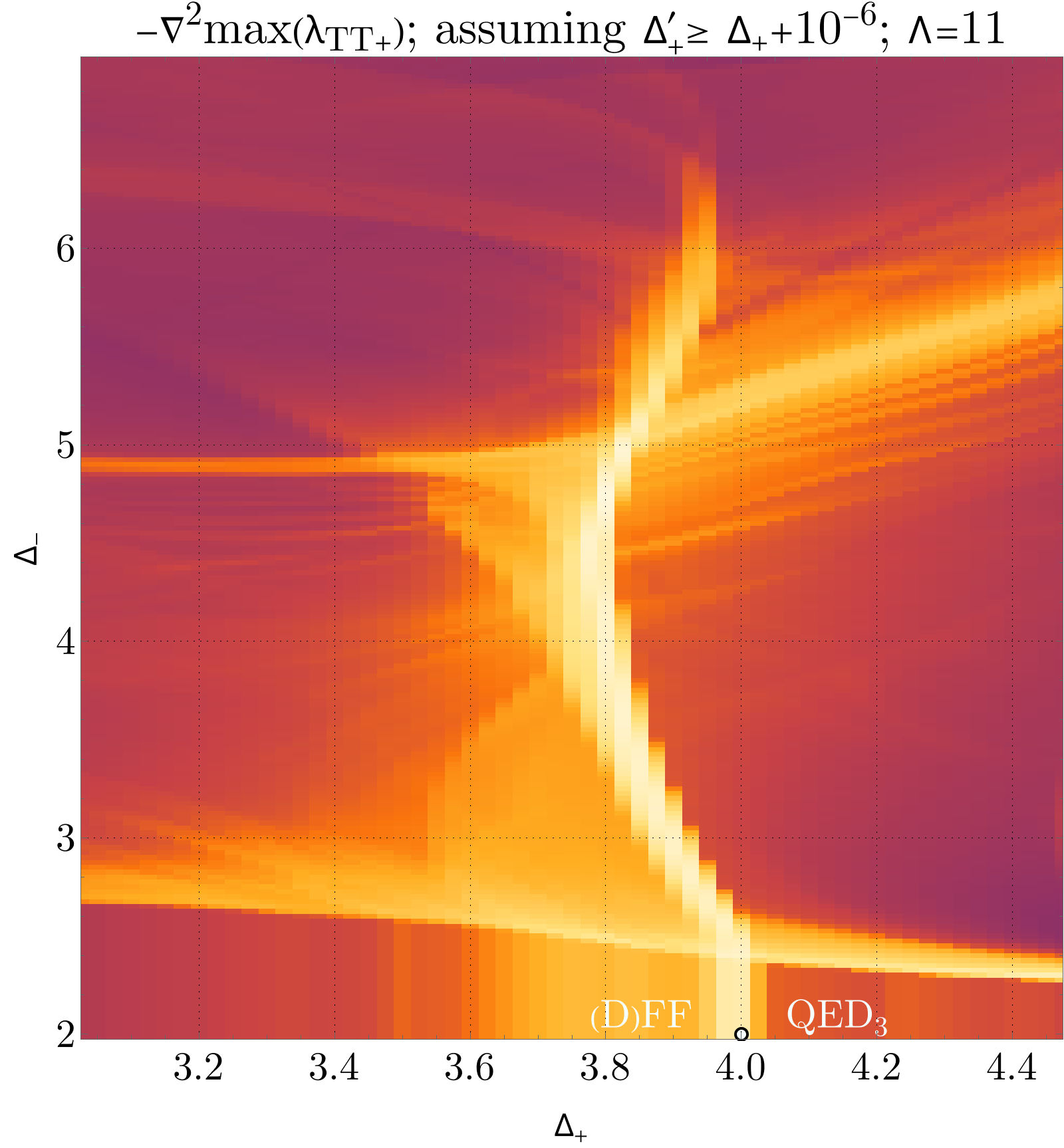}
  \caption{Higher-resolution Laplacian plot, showing the Dirac ridge in detail}
  \label{fig:max-lttp-dl-zoom2}
\end{figure}

In figure \ref{fig:max-lttp-dl-zoom1} we present a higher-resolution inset for the region \(\Delta_+\in[3.7,6.1]\), \(\Delta_-\in[2,7]\). The added detail shows there being almost organically curving features towards the rightmost extent of the plot extending downward from the MFT solution, almost like locks of hair. There is also an isolated curved feature potentially connecting to a vertical ridge around \(\Delta_+\simeq 4.7\). The diagonal ridge going toward the MFT solution reveals itself to be numerous such ridges in parallel, up to the limits of resolution. (That the ridge does not go directly to the MFT solution but to a slightly higher \(\Delta_-\) gap is an effect due to the low numerical order's upper bound on \(\Delta_-\) being 8; we will return to this point later.)

In figure \ref{fig:max-lttp-dl-zoom2} we show an even higher-resolution plot, an inset of the region \(\Delta_+\in[3,4.5]\), \(\Delta_-\in[2,7]\), focusing on the ``Dirac'' ridge as it intersects the ``Majorana'' ridge. This plot reveals a tremendous amount of rich substructure surrounding each of the most dramatic features. Below the strong horizontal line around \(\Delta_-=4.9\) there are numerous smaller and possibly more gently curved features. Below the large and diffuse diagonal line going off to the MFT solution are other parallel trajectories, as many as the resolution will allow us to see.

The natural question at this juncture is to ask whether these features are physically meaningful and/or in correspondence with any known CFTs. While the bottom of the ``Dirac'' ridge clearly has a physical interpretation, corresponding both to the Dirac free fermion theory with \((\Delta_+,\Delta_-)=(4,2)\) and to the large-\(N_f\) limit of conformal QED\(_3\). However, the exact correspondence here is potentially unclear; the fact that the ridge is vertical in the plot implies that the bootstrap solution has a leading parity-odd scalar a bit higher than the exact value of 2. It could be that this is a numerical convergence issue due to the low derivative order of this plot, or it could be that there is a different theory saturating the bound. 

Beyond this, it's difficult to construct local, unitary CFTs with perturbative Lagrangian descriptions that have the appropriate spectral gaps in this neighborhood. With a scalar theory, we can always construct parity-even singlet bilinears with leading order dimension 1; with a fermion theory, we can always construct parity-\emph{odd} singlet bilinears with leading order dimension 2. Leading dimensions of \((1,2)\) are exceptionally light in comparison to the upper extremities of our features which are around \((4,6)\). We note that with the assumption of a nonabelian current \(J^A\), it's possible to construct candidate descriptions for operators that could be the leading scalars:
\begin{align}
  \mathcal{O}_+^0 &\overset{?}{\equiv} J^2\\
  \mathcal{O}_-^0 &\overset{?}{\equiv} \star f_{ABC} (J^A\wedge J^B\wedge J^C)
\end{align}
and we can see that with some putative weakly coupled description these scaling dimensions would be at leading order 4 and 6, respectively. We also note that there could be a different parity odd operator
\begin{equation}
  \mathcal{O}_-^0 \overset{?}{\equiv} \star \delta_{AB} (J^A \wedge d J^B)
\end{equation}
which would have dimension 5. Could this correspond to the horizontal ridge we see at dimension 5? Indeed, these operators describe the leading scalars in the so-called generalized free vector field (GFVF) theory discussed in \cite{Dymarsky:2017xzb}. We should note that such a generalized free theory necessarily decouples from the stress tensor, so the GFVF is definitively \emph{not} the cause for the features we see. However, in an interacting theory, one might expect an analogous operator.

All of this discussion is moot if the features on display are unphysical in some way. Numerical bootstrap bounds often have spurious solutions that disappear under closer inspection. For example, in figure \ref{fig:tttt-allowed}, there is a seemingly nontrivial region in the far lower right with \(\Delta_+\geq 6\). Besides the fact that it seems to converge quickly with numerical order, that region is actually disallowed by further bootstrap constraints that assume that the leading parity even scalar is in the leading parity odd scalar's OPE i.e.\ \(\mathcal{O}_-\times\mathcal{O}_- \supset \mathcal{O}_+\) as discussed in \cite{Dymarsky:2017yzx}. While the best proof that features in bootstrap bounds correspond to real CFTs is identifying those CFTs through independent means, here we present some further plots to corroborate the claim that the features we find are nontrivial, sharply-defined, and not due to finite numerical order.

\begin{figure}
\centering
\begin{minipage}{.5\textwidth}
  \centering
  \includegraphics[width=0.97\linewidth]{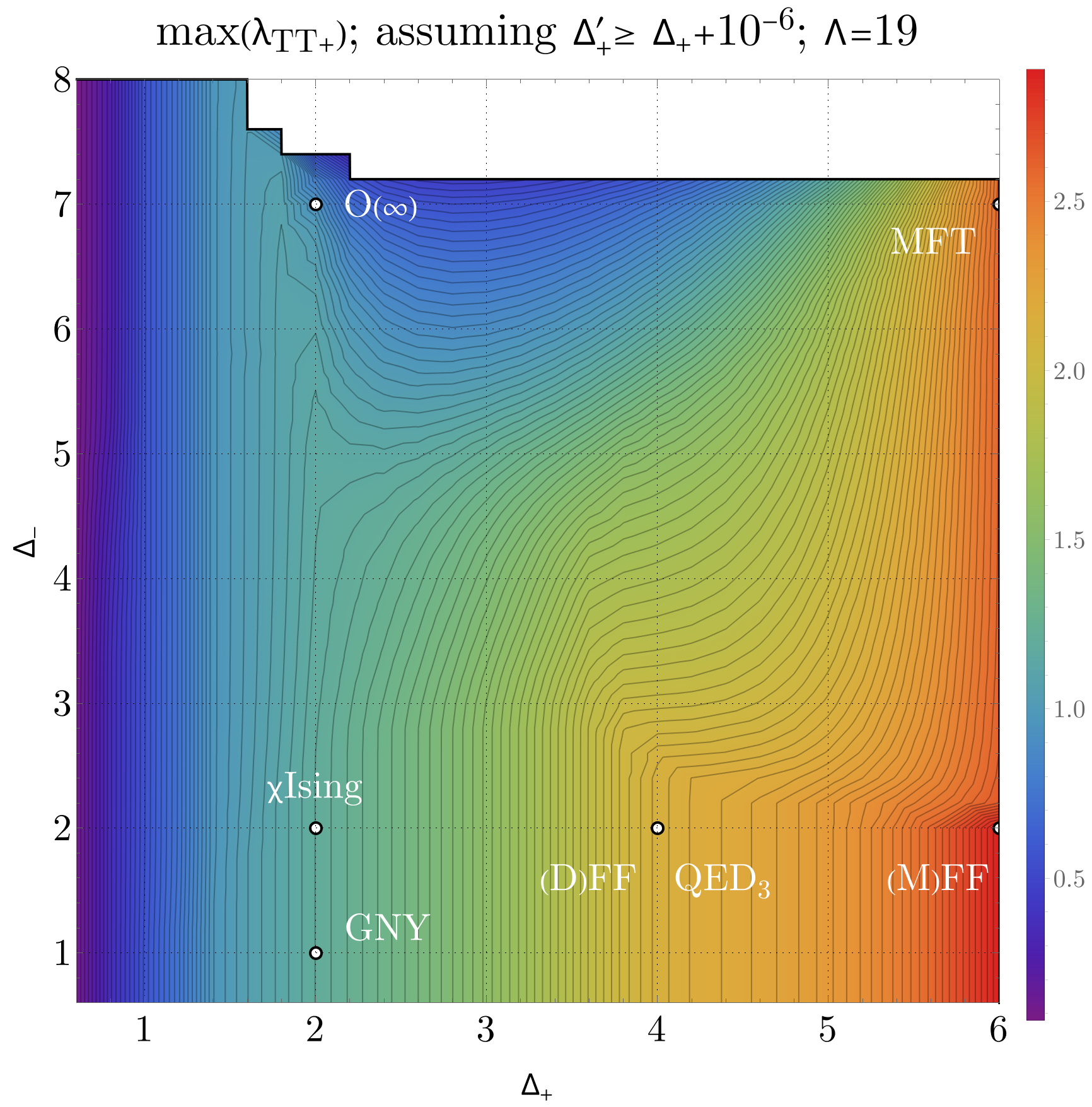}
\end{minipage}
\begin{minipage}{.5\textwidth}
  \centering
  \includegraphics[width=0.9\linewidth]{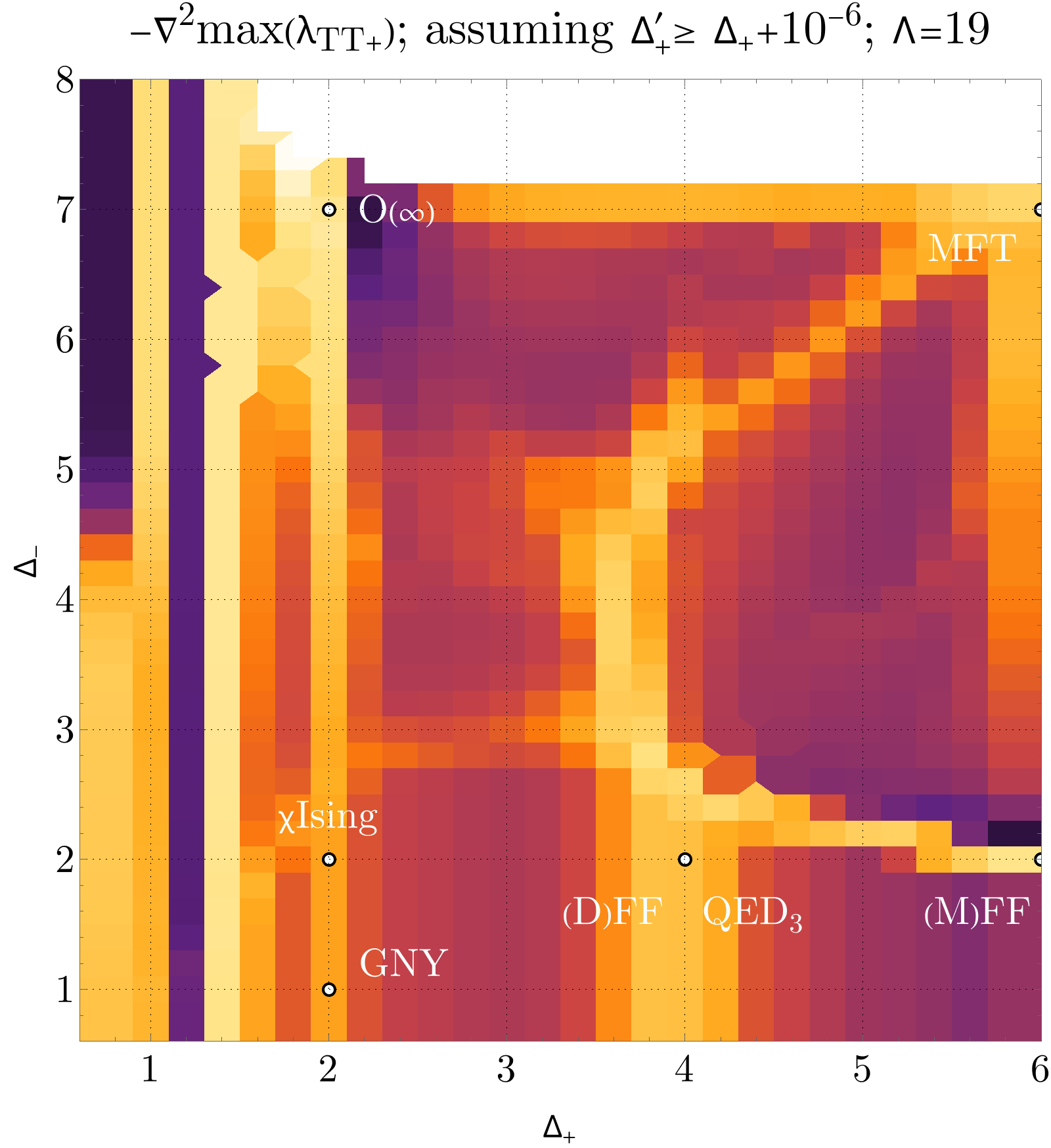}
\end{minipage}
\caption{Maximal $\lambda_{TT+}$ and discrete Laplacian (compare with figures   \ref{fig:max-lttp-contour} and \ref{fig:max-lttp-dl}) for the center of the allowed region, at $\Lambda = 19$. As we increased $\Lambda$, the MFT ridge appears to move closer to the line between $(4,5)$ and $(6,7)$.}
\label{fig:max-lttp-nmax10}
\end{figure}

First, we present the same OPE coefficient bound computed at higher derivative order, \(\Lambda=19\), presented in figure \ref{fig:max-lttp-nmax10}. As these computations were far more costly, the resolution is lower. (The grid is irregular due to some missing points from supercomputing cluster errors and not genuine physics; this irregularity leads to the plot itself having irregular cells.) Nonetheless, even at this low resolution we can see that the large scale features are not only persistent but fairly stable relative to their \(\Lambda=11\) appearance. The most dramatic change is that the diagonal ``MFT ridge'' has had its upper extent shifted down, matching with the convergence of the exclusion bounds themselves. We expect therefore that the sharp features we have been discussing are more than mere numerical artifacts.

\begin{figure}
  \centering
  \includegraphics[width=\textwidth]{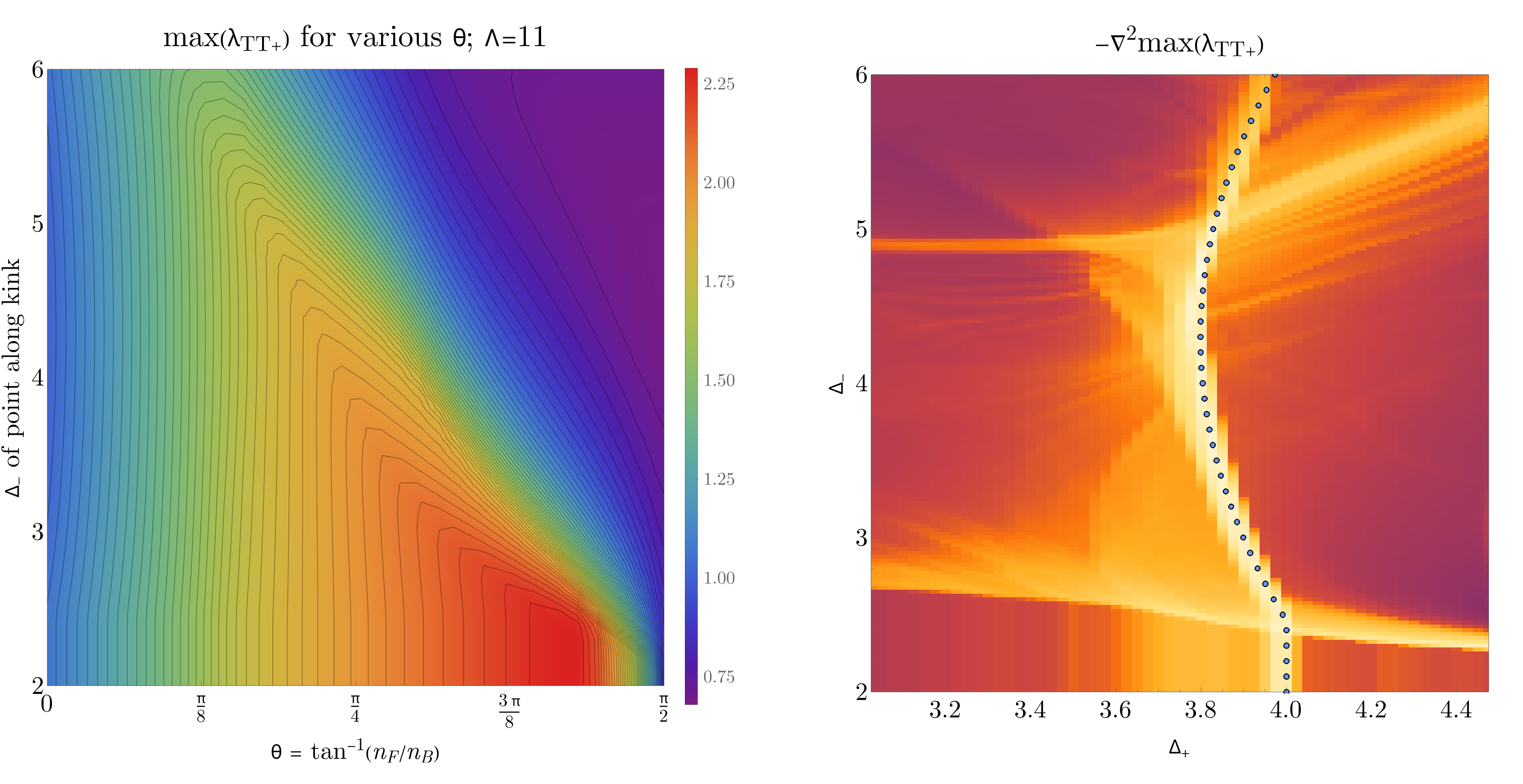}
  \caption{Maximal $\lambda_{TT+}$ as a function of $\theta$ (see text), for various points on the Dirac ridge. The saturating theory becomes increasingly bosonic as $\Delta_-$ increases, with $\theta_{\textrm{max}} \approx \pi/6$ when the Dirac ridge intersects the MFT ridge.}
  \label{fig:theta-contour}
\end{figure}

Next, returning to \(\Lambda=11\), we now show the OPE bound for points along the Dirac ridge assuming specific values of the parameter \(\theta\) (i.e.\ enforcing a unique stress tensor\footnote{As we don't impose a large gap above the stress tensor in the spin-2 parity-even sector, there is room for the sharing effect to weaken this assumption. See \cite{Liu:2020tpf} for related discussion on the sharing effect.}) rendered as a contour plot in figure \ref{fig:theta-contour}. As a reminder, this parameter is an angle in the space of \(\langle TTT\rangle\) coefficients where \(\theta=0\) for free scalar theories, \(\theta=\frac{\pi}{2}\) for free fermion theories, and any interacting theory must have intermediate values. We can see that throughout the Dirac ridge, the OPE bound has a fairly sharp peak as a function of \(\theta\). In the vicinity of the Dirac free fermion we see the maximum is close to but some small distance away from the free fermion value of \(\theta=\frac{\pi}{2}\). It could either be that the numerics have not yet converged to the free fermion value or that the theory saturating the bound is not the Dirac free fermion itself. Regardless, the sharp definition as a function of \(\theta\) means the bootstrap solutions require a finite \(c_T\) and a specific \(\langle TTT\rangle\) three point function, which we see as corroborating evidence toward there being real, nontrivial CFT solutions along the ridge. Curiously this ridge has a steady trajectory to lower and lower values of \(\theta\) as we increase the odd scalar gap. Crudely, one could describe this as saying the theories become ``more bosonic,'' though we should emphasize that it's unclear that there's any obvious, non-perturbative interpretation where this is meaningful.

In fact, due to a lack of next-to-leading-order perturbative results for \(\langle TTT\rangle\) entirely, it remains difficult to imagine what theories could correspond to the Dirac ridge. The sole perturbative results that we were able to find in the literature are to leading order for \(U(1)\) gauge theories with large \((N_f, N_s)\) fermions and scalars, respectively, reproduced in our notation \cite{Chowdhury:2012km}:
\begin{equation}
  \theta = \tan^{-1}\frac{N_f}{N_s} + O(N_f^{-1})+O(N_s^{-1}).
\end{equation}
This just reflects the UV matter content and doesn't yet incorporate any corrections from interaction. We can imagine that we can at least reproduce the ``sweep'' across \(\theta\) seen in our results by adjusting the matter content, but this is a somewhat vacuous point without more results.

As noted earlier however, the gaps to the leading scalars in this region already make the simplest Lagrangian descriptions impossible. It's of course possible that there is no single-parameter family of theories and this apparently sharp feature instead corresponds to pieces of many, many complicated theories that have nothing to do with each other. But we will posit strong-coupling limits of non-abelian conformal gauge theories as candidate Lagrangian descriptions.

\section{Other bounds: parity-odd scalars, assuming \(c_T\) and \(\theta\), current bootstrap}
\label{sec:other}

\begin{figure}
  \centering
  \includegraphics[width=0.6\textwidth]{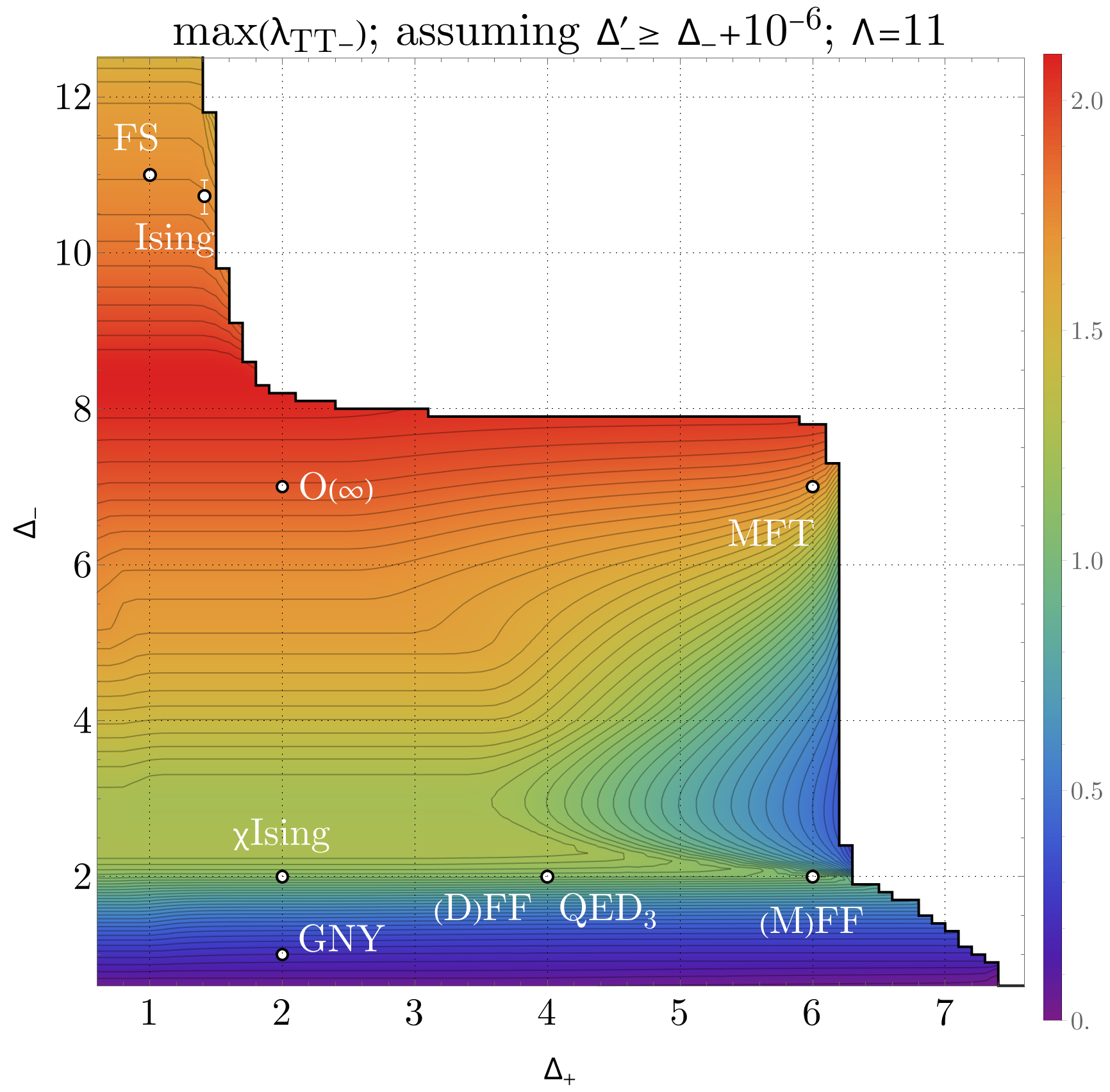}
  \caption{Contour plot, maximizing $\lambda_{TT-}$ across the entire allowed region. As in figure \ref{fig:max-lttp-contour}, we see kinks in the contours near identified CFTs.}
  \label{fig:max-lttm-contour}
\end{figure}

\begin{figure}[h]
  \centering
  \includegraphics[width=0.6\textwidth]{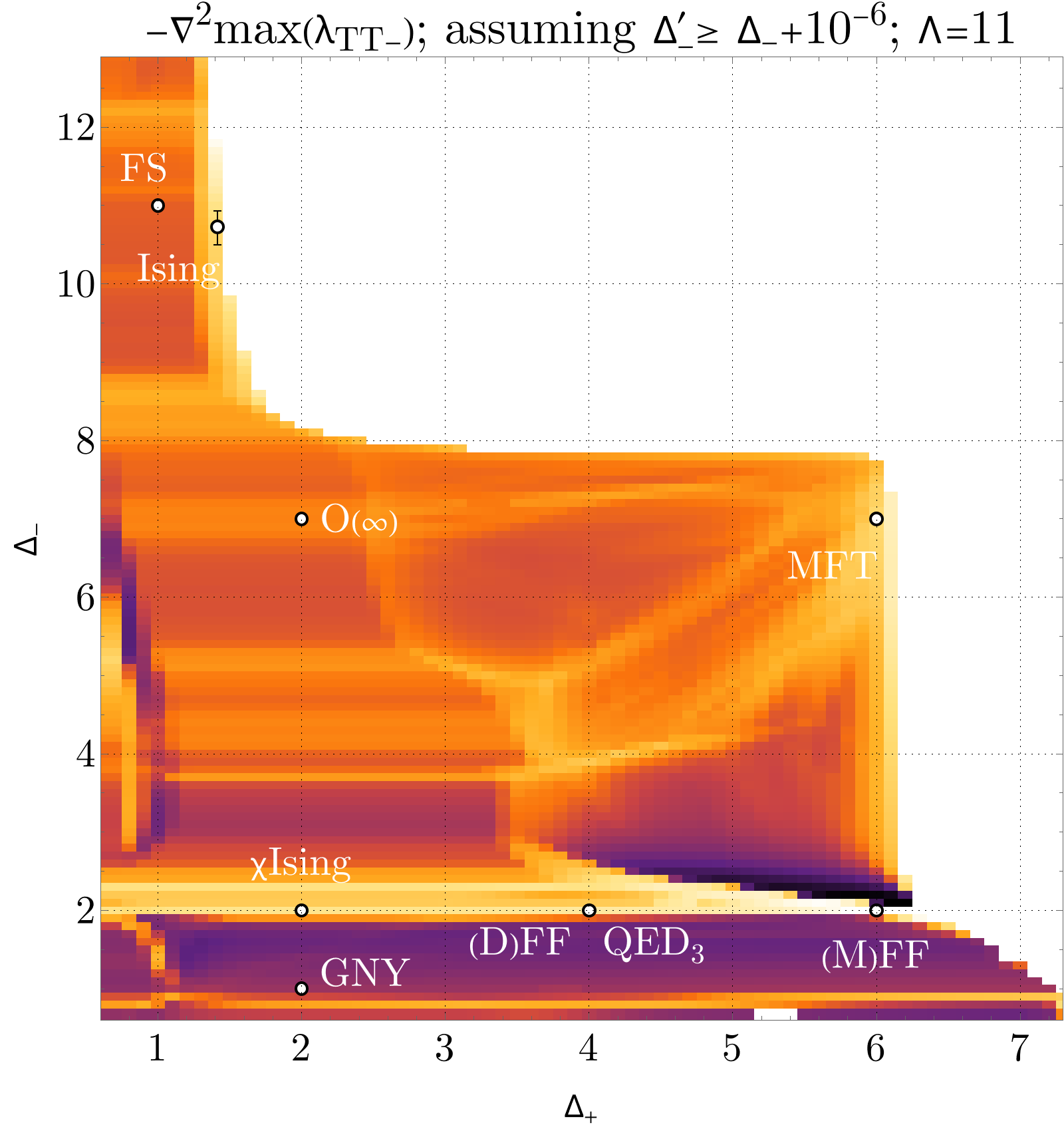}
  \caption{Discrete Laplacian of the maximal $\lambda_{TT-}$ (see contour plot in figure \ref{fig:max-lttm-contour}).}
  \label{fig:max-lttm-dl}
\end{figure}

We would be negligent if we did not admit that there is, in principle, nothing especially unique to the \(\lambda_{TT+}\) OPE coefficient or to the leading parity even scalar. We therefore briefly investigated the bounds derived from maximizing the leading parity-odd scalar OPE coefficient \(\lambda_{TT-}\). We present those results in figures \ref{fig:max-lttm-contour} and \ref{fig:max-lttm-dl} as a contour plot and a discrete Laplacian plot, respectively.

We can see that there are a completely different set of kinks, ridges, and other sharp features that appear. The straight horizontal lines are analogous to the straight vertical lines that appear for the \(\lambda_{TT+}\) bounds. As the corresponding axis is for a gap assumption, if that gap is not saturated the bound will not change in changing the gap, leading to these straight lines. Where these straight lines end is a lengthy ridge analogous to the Majorana ridge from the \(\lambda_{TT+}\) bounds: seeing as it originates in the vicinity of the free scalar theory in the upper right, we will refer to it as the ``free scalar ridge.'' Starting from that area, the ridge forms the boundary of the scalar gap exclusion plot where the \(O(N)\) archipelago lives, but upon getting to the \(O(\infty)\) vicinity it continues into the interior of the allowed area. From there it wends through a complicated region before finally apparently ending at the Majorana free fermion theory. Along the way it crosses a number of mysterious ridges in fairly complicated fashion in \(4\lesssim \Delta_-\lesssim 5\); there are two sharp ridges at \(\Delta_-=2\) (corresponding with the Majorana free fermion) and \(\Delta_-\approx 2.3\).

In the region of low \(\Delta_+\) on the left of the plot, there are apparently two nontrivial ridges, though the resolution available doesn't make it easy to discern what might be present. Likewise the region to the upper right seems to have a maelstrom of activity that's difficult to separate. We note that the nexus of that maelstrom is again the dimensions of \((\Delta_+,\Delta_-)\approx(4,5)\) that were puzzling in the even scalar coefficient plots. It's not obvious whether these complexities have anything to do with each other: we are extremizing different OPE coefficients, after all! But in either case this parameter space seems to contain a richness of structure that is far from being understood.

Finally, while not included in this present work, we have also made exploratory computations for similar OPE coefficient bounds (for the leading scalars) from the abelian current bootstrap and found similar structures present. Clearly there is a lot to do!

\section{Outlook: an atlas of 3D CFTs?}
\label{sec:outlook}

In this work we have shown nontrivial features of the allowed region of local, unitary 3D CFTs subject to constraints from the stress tensor four point function. While some features can be positively identified with known CFTs such as the scalar and fermion free theories but also the Ising CFT, there are many others which we are unable as yet to identify. Indeed, one could argue that absent some similar positive identification, it's unclear whether any of these features are even physical CFTs. They could merely be mathematical curiosities in the space of bootstrap solutions, or they could be bona fide field theories that nonetheless are too exotic or too complicated to be of interest to the study of field theories. We simply cannot say for certain! In any case it's not even \textit{a priori} a given that ``What is the bootstrap solution that maximizes the \(\lambda_{TT+}\) OPE coefficient?'' is a question that would yield anything of physical interest. But the fact that some features \emph{do} correspond very clearly with a known interacting CFT (Ising) buoys our confidence that there is something of physical relevance in our findings.

At this juncture, we would like to highlight how this work fits somewhat uneasily into the broader scientific program of CFTs. The gold standard for a conformal bootstrap study is to target some independently identified CFT so as to corroborate and massively improve upon existing data, all of which we don't do in the present work. CFTs are physically relevant as effective descriptions of the IR fixed points of Lagrangian QFT or lattice Hamiltonian models, after all, so it makes sense for the bootstrap to follow rather than lead. At the same time, there are CFTs which haven't been studied in great detail through non-bootstrap methods due to strong coupling or sign problems or other weaknesses of those methods which are nonetheless physically interesting. 

For example, conformal QED\(_3\) is understood chiefly through large-\(N_f\) perturbation theory, but its fate at small but finite \(N_f\) remains a mystery. Progress has been made in bootstrapping these theories \cite{Chester:2016wrc,Li:2018lyb,Albayrak:2021xtd,He:2023ewx} inspired by the perturbative results, but they have so far fallen short of replicating the successes of the \(O(N)\) and Gross-Neveu-Yukawa archipelago bootstraps for small \(N_f\). There are numerous challenges, but at least some may be summarized as deriving from the limitations of the Lagrangian QFT and lattice Hamiltonian models themselves: there is not enough meat on the bones for a bootstrap study to be able to rely solely on those results. For the technical experts, one key missing ingredient is a spectral gap derived from equations of motion which enforces a stronger form of locality: see Appendix \ref{app:locality} for more discussion.

It's in this context that the bootstrap could take some tentative steps toward leading rather than following. Instead of targeting e.g.\ \(N_f=4\) QED\(_3\), we could pursue a project of some flavor of CFT cartography. By this we mean mapping out a global atlas of CFTs in lieu of hunting for specific islands; perhaps we will stumble upon sharp features on that atlas which can then be identified with known CFTs in the future. Doing so with conserved currents (including the stress tensor) is the natural choice, not only because their correlation functions are guaranteed to be present in CFTs with their corresponding symmetries, but also because the stress tensor's presence is another strong locality constraint.

We should note that while we have significantly improved the numerical bootstrap's overall performance in nearly a decade of development, a single stress tensor bootstrap computation is still significantly more expensive than a scalar or fermion computation at the same numerical order. Additionally, the kinds of computations presented in this paper --- fine-mesh grids of optimization bounds --- multiplicatively increase the number of computations that need to be done relative to e.g. the mesh-triangulation based searches used in other studies. However, this need not mean that these computations are impractical! The results of this work were almost entirely found at low numerical order (\(\Lambda=11,19\)). Though they're more expensive for the reasons laid out, the lower numerical order counterbalances the overall cost relative to a more traditional computation at very high numerical order.

Whether or not the reader is convinced by this proposal, the depth of the conformal bootstrap remains alluring. It's remarkable that anything at all can be understood nonperturbatively about conformal field theories, and doubly so with such generic assumptions. How much the bootstrap knows about specific CFTs and how it is able to do so are questions that deserve ongoing investigation and research.

\section*{Acknowledgements}
We thank Cyuan-Han Chang, Vasiliy Dommes, Alexandre Homrich, Petr Kravchuk, Aike Liu, David Poland, David Simmons-Duffin, Petar Tadi\'{c}, Naveen Umasankar, Balt van Rees, Slava Rychkov, and Alessandro Vichi for useful discussions. RSE is supported by Simons Foundation grant 915279 (IHES).

Our computations were carried out at the following facilities:
\begin{itemize}
\item Expanse cluster at the San Diego Supercomputing Center (SDSC) through allocation PHY190023 from the Advanced Cyberinfrastructure Coordination Ecosystem: Services \& Support (ACCESS) program, which is supported by National Science Foundation grants \#2138259, \#2138286, \#2138307, \#2137603, and \#2138296. 
\item Yale Grace computing cluster, supported by the facilities and staff of the Yale University Faculty of Sciences High Performance Computing Center.
\end{itemize}

\newpage

\appendix

\section{Numerical parameters} \label{sec:params}

\begin{table}[h!]
  \centering
  \begin{tabular}{l c c c c c}
    \hline
    \hline
    \textbf{Order}                        &$\Lambda=11$ &$\Lambda=19$ &$\Lambda=27$ &$\Lambda=35$ &$\Lambda=43$ \\
    \hline
    \textbf{Parameters for \texttt{blocks\_3d}} & & & & & \\
    \texttt{keptPoleOrder}                & 8           & 14          & 32          & 106         & 130         \\
    \texttt{numPoles}                     & N/A         & N/A         & N/A         & 35          & 35          \\
    \texttt{order}                        & 60          & 60          & 80          & 106         & 130         \\
    \texttt{spins}                        & $S_{11}$    & $S_{19}$    & $S_{27}$    & $S_{35}$    & $S_{43}$    \\
    \texttt{precision}                    & 768         & 768         & 960         & 832         & 832         \\
    \hline
    \textbf{Parameters for \texttt{SDPB}} & & & & & \\
    \texttt{precision}                    & 768         & 768         & 960         & 832         & 832         \\
    \texttt{dualityGapThreshold}          & $10^{-30}$  & $10^{-30}$  & $10^{-30}$  & $10^{-30}$  & $10^{-75}$  \\
    \texttt{primalErrorThreshold}         & $10^{-200}$ & $10^{-200}$ & $10^{-200}$ & $10^{-200}$ & $10^{-200}$ \\
    \texttt{dualErrorThreshold}           & $10^{-200}$ & $10^{-200}$ & $10^{-200}$ & $10^{-200}$ & $10^{-200}$ \\
    \texttt{findPrimalFeasible}           & false       & false       & false       & false       & false       \\
    \texttt{findDualFeasible}             & false       & false       & false       & false       & false       \\
    \texttt{detectPrimalFeasibleJump}     & true        & true        & true        & true        & true        \\
    \texttt{detectDualFeasibleJump}       & true        & true        & true        & true        & true        \\
    \texttt{initialMatrixScalePrimal}     & $10^{20}$   & $10^{40}$   & $10^{50}$   & $10^{20}$   & $10^{20}$   \\
    \texttt{initialMatrixScaleDual}       & $10^{20}$   & $10^{40}$   & $10^{50}$   & $10^{20}$   & $10^{20}$   \\
    \texttt{feasibleCenteringParameter}   & 0.1         & 0.1         & 0.1         & 0.1         & 0.1         \\
    \texttt{infeasibleCenteringParameter} & 0.3         & 0.3         & 0.3         & 0.3         & 0.3         \\
    \texttt{stepLengthReduction}          & 0.7         & 0.7         & 0.7         & 0.7         & 0.7         \\
    \texttt{maxComplementarity}           & $10^{100}$  & $10^{100}$  & $10^{160}$  & $10^{160}$  & $10^{200}$  \\
    \texttt{minPrimalStep}                & 0           & 0           & 0           & 0           & 0           \\
    \texttt{minDualStep}                  & 0           & 0           & 0           & 0           & 0           \\
    \hline
    \hline
  \end{tabular}
  \caption{Numerical parameters used in feasibility searches (figure \ref{fig:tttt-allowed}). At $\Lambda=35$ and $\Lambda=43$, we computed blocks using optimal interpolation \cite{Chang:2025mwt} instead of the older pole-shifting method, hence the \texttt{numPoles} parameter.}
  \label{tab:params1}
\end{table}

\begin{table}[h!]
  \centering
  \begin{tabular}{l c c c}
    \hline
    \hline
    \textbf{Order}                        &$\Lambda=11$ &$\Lambda=19$ &$\Lambda=27$ \\
    \hline
    \textbf{Parameters for \texttt{blocks\_3d}} & & & \\
    \texttt{keptPoleOrder}                & 8           & 14          & 32          \\
    \texttt{order}                        & 60          & 60          & 80          \\
    \texttt{spins}                        & $S_{11}$    & $S_{19}$    & $S_{27}$    \\
    \texttt{precision}                    & 768         & 768         & 960         \\
    \hline
    \textbf{Parameters for \texttt{SDPB}} & & & \\
    \texttt{precision}                    & 960--1024*  & 960         & 1024        \\
    \texttt{dualityGapThreshold}          & $10^{-30}$  & $10^{-30}$  & $10^{-30}$  \\
    \texttt{primalErrorThreshold}         & $10^{-30}$  & $10^{-30}$  & $10^{-40}$  \\
    \texttt{dualErrorThreshold}           & $10^{-30}$  & $10^{-30}$  & $10^{-40}$  \\
    \texttt{findPrimalFeasible}           & false       & false       & false       \\
    \texttt{findDualFeasible}             & false       & false       & false       \\
    \texttt{detectPrimalFeasibleJump}     & false       & false       & false       \\
    \texttt{detectDualFeasibleJump}       & false       & false       & false       \\
    \texttt{initialMatrixScalePrimal}     & $10^{20}$   & $10^{40}$   & $10^{50}$   \\
    \texttt{initialMatrixScaleDual}       & $10^{20}$   & $10^{40}$   & $10^{50}$   \\
    \texttt{feasibleCenteringParameter}   & 0.1         & 0.1         & 0.1         \\
    \texttt{infeasibleCenteringParameter} & 0.3         & 0.3         & 0.3         \\
    \texttt{stepLengthReduction}          & 0.7         & 0.7         & 0.7         \\
    \texttt{maxComplementarity}           & $10^{100}$  & $10^{100}$  & $10^{160}$  \\
    \texttt{minPrimalStep}                & 0           & 0           & 0           \\
    \texttt{minDualStep}                  & 0           & 0           & 0           \\
    \hline
    \hline
  \end{tabular}  
  \caption{Numerical parameters used in extremization searches (figures \ref{fig:max-lttp-side-by-side}-\ref{fig:max-lttm-dl}). At $\Lambda = 11$, we used a solver precision of 960 in every case \emph{except} the scan over $\theta$, where a precision of 1024 was required for several points.}
  \label{tab:params2}
\end{table}

Table \ref{tab:params1} and table \ref{tab:params2} illustrate numerical parameters used for the feasibility and optimization searches respectively. The sets of spins used at each derivative order are:
\begin{align*}
  S_{11} &= \{0, 0.5, 1 \dots 21.5\} \\
  S_{19} &= \{0, 0.5, 1 \dots 26.5\} \cup \{49,49.5,50,50.5\} \\
  S_{27} &= \{0, 0.5, 1 \dots 44.5\} \cup \{47 \dots 48.5\} \cup \{51 \dots 52.5\}
           \cup \{55 \dots 56.5\}\\
  &\qquad\cup \{59 \dots 60.5\} \cup \{63 \dots 64.5\} \cup \{67 \dots 68.5\} \\
  S_{35} &= \{0, 1  , 2 \dots 75\} \\
  S_{43} &= \{0, 1  , 2 \dots 90\}.
\end{align*}

Note that, with the exception of the $\Lambda=35$ and $43$ scalar exclusion runs, our computations did not use the new versions of \texttt{SDPB} and \texttt{blocks\_3d} developed in \cite{Chang:2024whx,Chang:2025mwt}. Reproducing our results with state-of-the-art numerics would likely require lower precision at a given order.

\section{On locality}
\label{app:locality}
Here we will briefly comment on varying definitions and notions of locality used in the context of CFT and the conformal bootstrap. Roughly speaking there are two, in the sense of cutting and gluing and in the sense of the existence of the stress tensor. The distinction is important as there exist theories that satisfy the former but do not satisfy the latter, and various bootstrap assumptions only require the former.

Indeed, the core assumption of the conformal bootstrap, the existence of the OPE, requires cutting-and-gluing locality \cite{Moore:2025tmt} by way of the state-operator correspondence \cite{Simmons-Duffin:2016gjk,Poland:2018epd}. That is, one must be able to evaluate path integrals on slices of spacetime to prepare states on those slices' boundaries, and in turn be able to glue those slices together (with inner products on the state spaces) to evaluate the full spacetime's path integral. In this sense no conformal bootstrap result is truly nonlocal: there is always a sense of locality in the path integral.

However, the conformal bootstrap does \emph{not} require the existence of the stress tensor. Mechanically, there is nothing preventing the exchanged stress tensor from having vanishing OPE coefficients. Indeed, bootstrap bounds are generically satisfied by generalized free theories (GFFs) which have nonlocal propagators and lack a conserved stress tensor, a feature used in the Navigator Function \cite{Reehorst:2021ykw} to improve numerical search routines in bootstrap computations. There are also CFTs with long-range interactions such as the long-range Ising (LRI) CFT \cite{Fisher:1972zz,Paulos:2015jfa,Behan:2017emf,Behan:2018hfx,Behan:2023ile} without conserved stress tensors, and this is also a generic feature of nontrivial boundary and defect CFTs \cite{Billo:2016cpy,DiPietro:2019hqe}. These CFTs are local ``enough'' to have perfectly healthy OPEs, but they are still ``less'' local than those with stress tensors.

Indeed, one can argue that much of the magic of the conformal bootstrap comes from being able to enhance the locality condition on the CFT from merely cutting-and-gluing. In all of the cases where bootstrap ``islands'' have been found, a gap assumption on the spectrum inspired by a Lagrangian equation of motion has been necessary to exclude GFF and other nonlocal solutions \cite{Kos:2014bka,Kos:2015mba,Rong:2018okz,Atanasov:2018kqw,Erramilli:2022kgp}; the multiplet recombination implied by such equations of motion also relate to the conservation of the stress tensor \cite{Behan:2017emf} (which itself can be interpreted as an equation of motion). It seems there are many OPEs, but only a select few are local in this strong sense.

Nonetheless, the class of CFTs with Lagrangian UV descriptions with equations of motion that provide useful spectral gap assumptions is fairly narrow and excludes conformal gauge theories such as conformal QED\(_3\). Indeed, cQED\(_3\) appears to have a nonlocal marginal deformation \cite{DiPietro:2019hqe,Heydeman:2020ijz}; this nonlocal conformal manifold pollutes bootstrap bounds of the monopole sectors \cite{Albayrak:2021xtd}, and beyond conservation laws there are no gauge-invariant equations of motion that are useful at constraining the monopole spectrum and eliminating these nonlocal solutions.

How, then, to proceed? We can conjecture that one approach may be to enforce the existence of the stress tensor the one way we can, by including stress tensors as external operators in our constraints. It's not obvious that such an approach would work or to what efficacy, but given the seeming importance of locality, it would be worth trying.

\newpage

\small
\bibliographystyle{utphys}
\bibliography{mybib.bib}

@article{Erramilli:2019njx,
  author        = {Erramilli, Rajeev S. and Iliesiu, Luca V. and Kravchuk, Petr},
  title         = {{Recursion relation for general 3d blocks}},
  eprint        = {1907.11247},
  archiveprefix = {arXiv},
  primaryclass  = {hep-th},
  reportnumber  = {PUPT-2593},
  doi           = {10.1007/JHEP12(2019)116},
  journal       = {JHEP},
  volume        = {12},
  pages         = {116},
  year          = {2019}
}

@article{Erramilli:2020rlr,
  author        = {Erramilli, Rajeev S. and Iliesiu, Luca V. and Kravchuk, Petr and Landry, Walter and Poland, David and Simmons-Duffin, David},
  title         = {{blocks\_3d: software for general 3d conformal blocks}},
  eprint        = {2011.01959},
  archiveprefix = {arXiv},
  primaryclass  = {hep-th},
  reportnumber  = {CALT-TH 2020-048},
  doi           = {10.1007/JHEP11(2021)006},
  journal       = {JHEP},
  volume        = {11},
  pages         = {006},
  year          = {2021}
}

@article{Dymarsky:2017yzx,
  author        = {Dymarsky, Anatoly and Kos, Filip and Kravchuk, Petr and Poland, David and Simmons-Duffin, David},
  title         = {{The 3d Stress-Tensor Bootstrap}},
  eprint        = {1708.05718},
  archiveprefix = {arXiv},
  primaryclass  = {hep-th},
  reportnumber  = {CALT-TH-2017-043},
  doi           = {10.1007/JHEP02(2018)164},
  journal       = {JHEP},
  volume        = {02},
  pages         = {164},
  year          = {2018}
}

@article{Kos:2014bka,
  author        = {Kos, Filip and Poland, David and Simmons-Duffin, David},
  title         = {{Bootstrapping Mixed Correlators in the 3D Ising Model}},
  eprint        = {1406.4858},
  archiveprefix = {arXiv},
  primaryclass  = {hep-th},
  doi           = {10.1007/JHEP11(2014)109},
  journal       = {JHEP},
  volume        = {11},
  pages         = {109},
  year          = {2014}
}

@article{Kos:2016ysd,
  author        = {Kos, Filip and Poland, David and Simmons-Duffin, David and Vichi, Alessandro},
  title         = {{Precision Islands in the Ising and $O(N)$ Models}},
  eprint        = {1603.04436},
  archiveprefix = {arXiv},
  primaryclass  = {hep-th},
  reportnumber  = {CERN-TH-2016-050},
  doi           = {10.1007/JHEP08(2016)036},
  journal       = {JHEP},
  volume        = {08},
  pages         = {036},
  year          = {2016}
}

@article{Poland:2018epd,
  author        = {Poland, David and Rychkov, Slava and Vichi, Alessandro},
  title         = {{The Conformal Bootstrap: Theory, Numerical Techniques, and Applications}},
  eprint        = {1805.04405},
  archiveprefix = {arXiv},
  primaryclass  = {hep-th},
  doi           = {10.1103/RevModPhys.91.015002},
  journal       = {Rev. Mod. Phys.},
  volume        = {91},
  pages         = {015002},
  year          = {2019}
}

@article{Rychkov:2023wsd,
  author        = {Rychkov, Slava and Su, Ning},
  title         = {{New Developments in the Numerical Conformal Bootstrap}},
  eprint        = {2311.15844},
  archiveprefix = {arXiv},
  primaryclass  = {hep-th},
  month         = {11},
  year          = {2023}
}

@article{Dymarsky:2017xzb,
  author        = {Dymarsky, Anatoly and Penedones, Joao and Trevisani, Emilio and Vichi, Alessandro},
  title         = {{Charting the space of 3D CFTs with a continuous global symmetry}},
  eprint        = {1705.04278},
  archiveprefix = {arXiv},
  primaryclass  = {hep-th},
  doi           = {10.1007/JHEP05(2019)098},
  journal       = {JHEP},
  volume        = {05},
  pages         = {098},
  year          = {2019}
}

@article{Reehorst:2019pzi,
  author        = {Reehorst, Marten and Trevisani, Emilio and Vichi, Alessandro},
  title         = {{Mixed Scalar-Current bootstrap in three dimensions}},
  eprint        = {1911.05747},
  archiveprefix = {arXiv},
  primaryclass  = {hep-th},
  doi           = {10.1007/JHEP12(2020)156},
  journal       = {JHEP},
  volume        = {12},
  pages         = {156},
  year          = {2020}
}

@article{He:2023ewx,
  author        = {He, Yin-Chen and Rong, Junchen and Su, Ning and Vichi, Alessandro},
  title         = {{Non-Abelian currents bootstrap}},
  eprint        = {2302.11585},
  archiveprefix = {arXiv},
  primaryclass  = {hep-th},
  doi           = {10.1007/JHEP03(2024)175},
  journal       = {JHEP},
  volume        = {03},
  pages         = {175},
  year          = {2024}
}

@article{Mitchell:2024hix,
  author        = {Mitchell, Matthew S. and Poland, David},
  title         = {{Bounding irrelevant operators in the 3d Gross-Neveu-Yukawa CFTs}},
  eprint        = {2406.12974},
  archiveprefix = {arXiv},
  primaryclass  = {hep-th},
  doi           = {10.1007/JHEP09(2024)134},
  journal       = {JHEP},
  volume        = {09},
  pages         = {134},
  year          = {2024}
}

@article{Reehorst:2021ykw,
  author        = {Reehorst, Marten and Rychkov, Slava and Simmons-Duffin, David and Sirois, Benoit and Su, Ning and van Rees, Balt},
  title         = {{Navigator Function for the Conformal Bootstrap}},
  eprint        = {2104.09518},
  archiveprefix = {arXiv},
  primaryclass  = {hep-th},
  reportnumber  = {CPHT-RR032.042021},
  doi           = {10.21468/SciPostPhys.11.3.072},
  journal       = {SciPost Phys.},
  volume        = {11},
  pages         = {072},
  year          = {2021}
}

@article{Chang:2025mwt,
  author        = {Chang, Cyuan-Han and Dommes, Vasiliy and Kravchuk, Petr and Poland, David and Simmons-Duffin, David},
  title         = {{Accurate bootstrap bounds from optimal interpolation}},
  eprint        = {2509.14307},
  archiveprefix = {arXiv},
  primaryclass  = {hep-th},
  reportnumber  = {CALT-TH 2025-031},
  month         = {9},
  year          = {2025}
}

@article{Behan:2017emf,
  author        = {Behan, Connor and Rastelli, Leonardo and Rychkov, Slava and Zan, Bernardo},
  title         = {{A scaling theory for the long-range to short-range crossover and an infrared duality}},
  eprint        = {1703.05325},
  archiveprefix = {arXiv},
  primaryclass  = {hep-th},
  reportnumber  = {CERN-TH-2017-052, YITP-SB-17-13, CERN-PH-TH-2017-052},
  doi           = {10.1088/1751-8121/aa8099},
  journal       = {J. Phys. A},
  volume        = {50},
  number        = {35},
  pages         = {354002},
  year          = {2017}
}

@article{Liu:2020tpf,
  author        = {Liu, Junyu and Meltzer, David and Poland, David and Simmons-Duffin, David},
  title         = {{The Lorentzian inversion formula and the spectrum of the 3d O(2) CFT}},
  eprint        = {2007.07914},
  archiveprefix = {arXiv},
  primaryclass  = {hep-th},
  doi           = {10.1007/JHEP09(2020)115},
  journal       = {JHEP},
  volume        = {09},
  pages         = {115},
  year          = {2020},
  note          = {[Erratum: JHEP 01, 206 (2021)]}
}

@article{Chester:2019ifh,
  author        = {Chester, Shai M. and Landry, Walter and Liu, Junyu and Poland, David and Simmons-Duffin, David and Su, Ning and Vichi, Alessandro},
  title         = {{Carving out OPE space and precise $O(2)$ model critical exponents}},
  eprint        = {1912.03324},
  archiveprefix = {arXiv},
  primaryclass  = {hep-th},
  reportnumber  = {CALT-TH-2019-051},
  doi           = {10.1007/JHEP06(2020)142},
  journal       = {JHEP},
  volume        = {06},
  pages         = {142},
  year          = {2020}
}

@article{Chester:2020iyt,
  author        = {Chester, Shai M. and Landry, Walter and Liu, Junyu and Poland, David and Simmons-Duffin, David and Su, Ning and Vichi, Alessandro},
  title         = {{Bootstrapping Heisenberg magnets and their cubic instability}},
  eprint        = {2011.14647},
  archiveprefix = {arXiv},
  primaryclass  = {hep-th},
  reportnumber  = {CALT-TH-2020-053},
  doi           = {10.1103/PhysRevD.104.105013},
  journal       = {Phys. Rev. D},
  volume        = {104},
  number        = {10},
  pages         = {105013},
  year          = {2021}
}

@article{Erramilli:2022kgp,
  author        = {Erramilli, Rajeev S. and Iliesiu, Luca V. and Kravchuk, Petr and Liu, Aike and Poland, David and Simmons-Duffin, David},
  title         = {{The Gross-Neveu-Yukawa archipelago}},
  eprint        = {2210.02492},
  archiveprefix = {arXiv},
  primaryclass  = {hep-th},
  reportnumber  = {CALT-TH 2022-027},
  doi           = {10.1007/JHEP02(2023)036},
  journal       = {JHEP},
  volume        = {02},
  pages         = {036},
  year          = {2023}
}

@article{Atanasov:2022bpi,
  author        = {Atanasov, Alexander and Hillman, Aaron and Poland, David and Rong, Junchen and Su, Ning},
  title         = {{Precision bootstrap for the $ \mathcal{N} $ = 1 super-Ising model}},
  eprint        = {2201.02206},
  archiveprefix = {arXiv},
  primaryclass  = {hep-th},
  doi           = {10.1007/JHEP08(2022)136},
  journal       = {JHEP},
  volume        = {08},
  pages         = {136},
  year          = {2022}
}

@article{Chang:2024whx,
  author        = {Chang, Cyuan-Han and Dommes, Vasiliy and Erramilli, Rajeev S. and Homrich, Alexandre and Kravchuk, Petr and Liu, Aike and Mitchell, Matthew S. and Poland, David and Simmons-Duffin, David},
  title         = {{Bootstrapping the 3d Ising stress tensor}},
  eprint        = {2411.15300},
  archiveprefix = {arXiv},
  primaryclass  = {hep-th},
  reportnumber  = {CALT-TH 2024-047},
  doi           = {10.1007/JHEP03(2025)136},
  journal       = {JHEP},
  volume        = {03},
  pages         = {136},
  year          = {2025}
}

@article{DiPietro:2015taa,
  author        = {Di Pietro, Lorenzo and Komargodski, Zohar and Shamir, Itamar and Stamou, Emmanuel},
  title         = {{Quantum Electrodynamics in d=3 from the {\ensuremath{\varepsilon}} Expansion}},
  eprint        = {1508.06278},
  archiveprefix = {arXiv},
  primaryclass  = {hep-th},
  doi           = {10.1103/PhysRevLett.116.131601},
  journal       = {Phys. Rev. Lett.},
  volume        = {116},
  number        = {13},
  pages         = {131601},
  year          = {2016}
}

@article{Mitchell:2025oao,
  author        = {Mitchell, Matthew S. and Poland, David},
  title         = {{Classifying GNY-like models}},
  eprint        = {2512.11963},
  archiveprefix = {arXiv},
  primaryclass  = {hep-th},
  month         = {12},
  year          = {2025}
}

@article{Albayrak:2021xtd,
  author        = {Albayrak, Soner and Erramilli, Rajeev S. and Li, Zhijin and Poland, David and Xin, Yuan},
  title         = {{Bootstrapping $N_f$=4 conformal QED$_3$}},
  eprint        = {2112.02106},
  archiveprefix = {arXiv},
  primaryclass  = {hep-th},
  doi           = {10.1103/PhysRevD.105.085008},
  journal       = {Phys. Rev. D},
  volume        = {105},
  number        = {8},
  pages         = {085008},
  year          = {2022}
}

@article{fei2016yukawa,
  author        = {Fei, Lin and Giombi, Simone and Klebanov, Igor R. and
                   Tarnopolsky, Grigory},
  title         = {{Yukawa CFTs and Emergent Supersymmetry}},
  journal       = {PTEP},
  volume        = {2016},
  year          = {2016},
  number        = {12},
  pages         = {12C105},
  doi           = {10.1093/ptep/ptw120},
  eprint        = {1607.05316},
  archiveprefix = {arXiv},
  primaryclass  = {hep-th},
  reportnumber  = {PUPT-2504},
  slaccitation  = {%%CITATION = ARXIV:1607.05316;%%}
}

@article{vojta2000quantum,
  author       = {Vojta, Matthias and Zhang, Ying and Sachdev, Subir},
  title        = {{Quantum Phase Transitions in d-Wave Superconductors}},
  journal      = {Phys. Rev. Lett.},
  volume       = {85},
  year         = {2000},
  pages        = {4940-4943},
  doi          = {10.1103/PhysRevLett.85.4940},
  slaccitation = {%%CITATION = PRLTA,85,4940;%%}
}

@article{vojta2003quantum,
  title     = {Quantum phase transitions},
  author    = {Vojta, Matthias},
  journal   = {Reports on Progress in Physics},
  volume    = {66},
  number    = {12},
  pages     = {2069},
  year      = {2003},
  publisher = {IOP Publishing}
}

@article{Moon:2012rx,
  author        = {Moon, Eun-Gook and Xu, Cenke and Kim, Yong Baek and
                   Balents, Leon},
  title         = {{Non-Fermi liquid and topological states with strong
                   spin-orbit coupling}},
  journal       = {Phys. Rev. Lett.},
  volume        = {111},
  year          = {2013},
  pages         = {206401},
  doi           = {10.1103/PhysRevLett.111.206401},
  eprint        = {1212.1168},
  archiveprefix = {arXiv},
  primaryclass  = {cond-mat.str-el},
  slaccitation  = {%%CITATION = ARXIV:1212.1168;%%}
}

@article{Herbut:2014lfa,
  author        = {Herbut, Igor F. and Janssen, Lukas},
  title         = {{Topological Mott insulator in three-dimensional systems
                   with quadratic band touching}},
  journal       = {Phys. Rev. Lett.},
  volume        = {113},
  year          = {2014},
  pages         = {106401},
  doi           = {10.1103/PhysRevLett.113.106401},
  eprint        = {1404.5721},
  archiveprefix = {arXiv},
  primaryclass  = {cond-mat.str-el},
  slaccitation  = {%%CITATION = ARXIV:1404.5721;%%}
}

@article{Mihaila:2017ble,
  author        = {Mihaila, Luminita N. and Zerf, Nikolai and Ihrig, Bernhard and Herbut, Igor F. and Scherer, Michael M.},
  title         = {{Gross-Neveu-Yukawa model at three loops and Ising critical behavior of Dirac systems}},
  eprint        = {1703.08801},
  archiveprefix = {arXiv},
  primaryclass  = {cond-mat.str-el},
  doi           = {10.1103/PhysRevB.96.165133},
  journal       = {Phys. Rev. B},
  volume        = {96},
  number        = {16},
  pages         = {165133},
  year          = {2017}
}

@article{Zerf:2017zqi,
  author        = {Zerf, Nikolai and Mihaila, Luminita N. and Marquard, Peter and Herbut, Igor F. and Scherer, Michael M.},
  title         = {{Four-loop critical exponents for the Gross-Neveu-Yukawa models}},
  eprint        = {1709.05057},
  archiveprefix = {arXiv},
  primaryclass  = {hep-th},
  reportnumber  = {DESY-17-133},
  doi           = {10.1103/PhysRevD.96.096010},
  journal       = {Phys. Rev. D},
  volume        = {96},
  number        = {9},
  pages         = {096010},
  year          = {2017}
}

@article{Rosenstein:1993zf,
  author       = {Rosenstein, B. and Yu, Hoi-Lai and Kovner, A.},
  title        = {{Critical exponents of new universality classes}},
  reportnumber = {IP-ASTP-14-93, LA-UR-93-2562},
  doi          = {10.1016/0370-2693(93)91253-J},
  journal      = {Phys. Lett. B},
  volume       = {314},
  pages        = {381--386},
  year         = {1993}
}

@article{Gracey:1990wi,
  author       = {Gracey, J. A.},
  title        = {{Calculation of exponent eta to O(1/N**2) in the O(N) Gross-Neveu model}},
  reportnumber = {PRINT-90-0395 (HELSINKI), HU-TFT-90-48},
  doi          = {10.1142/S0217751X91000241},
  journal      = {Int. J. Mod. Phys. A},
  volume       = {6},
  pages        = {395--408},
  year         = {1991},
  note         = {[Erratum: Int.J.Mod.Phys.A 6, 2755 (1991)]}
}

@article{Chowdhury:2012km,
  author        = {Chowdhury, Debanjan and Raju, Suvrat and Sachdev, Subir and Singh, Ajay and Strack, Philipp},
  title         = {{Multipoint correlators of conformal field theories: implications for quantum critical transport}},
  eprint        = {1210.5247},
  archiveprefix = {arXiv},
  primaryclass  = {cond-mat.str-el},
  reportnumber  = {ICTS-2012-07, HRI-ST-1206},
  doi           = {10.1103/PhysRevB.87.085138},
  journal       = {Phys. Rev. B},
  volume        = {87},
  number        = {8},
  pages         = {085138},
  year          = {2013}
}

@article{Maldacena:2011jn,
  author        = {Maldacena, Juan and Zhiboedov, Alexander},
  title         = {{Constraining Conformal Field Theories with A Higher Spin Symmetry}},
  eprint        = {1112.1016},
  archiveprefix = {arXiv},
  primaryclass  = {hep-th},
  doi           = {10.1088/1751-8113/46/21/214011},
  journal       = {J. Phys. A},
  volume        = {46},
  pages         = {214011},
  year          = {2013}
}

@article{Meltzer:2018tnm,
  author        = {Meltzer, David},
  title         = {{Higher Spin ANEC and the Space of CFTs}},
  eprint        = {1811.01913},
  archiveprefix = {arXiv},
  primaryclass  = {hep-th},
  doi           = {10.1007/JHEP07(2019)001},
  journal       = {JHEP},
  volume        = {07},
  pages         = {001},
  year          = {2019}
}

@inproceedings{Simmons-Duffin:2016gjk,
  author        = {Simmons-Duffin, David},
  title         = {{The Conformal Bootstrap}},
  booktitle     = {{Theoretical Advanced Study Institute in Elementary Particle Physics}: {New Frontiers in Fields and Strings}},
  eprint        = {1602.07982},
  archiveprefix = {arXiv},
  primaryclass  = {hep-th},
  doi           = {10.1142/9789813149441_0001},
  pages         = {1--74},
  year          = {2017}
}

@inproceedings{Moore:2025tmt,
  author        = {Moore, Gregory W. and Saxena, Vivek and Freed, with an appendix by Daniel S.},
  title         = {{TASI Lectures On Topological Field Theories And Differential Cohomology}},
  booktitle     = {{Theoretical Advanced Study Institute in Elementary Particle Physics 2023}: {Aspects of Symmetry}},
  eprint        = {2510.07408},
  archiveprefix = {arXiv},
  primaryclass  = {hep-th},
  reportnumber  = {YITP-SB-2025-04},
  month         = {10},
  year          = {2025}
}

@article{Behan:2023ile,
  author        = {Behan, Connor and Lauria, Edoardo and Nocchi, Maria and van Vliet, Philine},
  title         = {{Analytic and numerical bootstrap for the long-range Ising model}},
  eprint        = {2311.02742},
  archiveprefix = {arXiv},
  primaryclass  = {hep-th},
  doi           = {10.1007/JHEP03(2024)136},
  journal       = {JHEP},
  volume        = {03},
  pages         = {136},
  year          = {2024}
}

@article{Behan:2018hfx,
  author        = {Behan, Connor},
  title         = {{Bootstrapping the long-range Ising model in three dimensions}},
  eprint        = {1810.07199},
  archiveprefix = {arXiv},
  primaryclass  = {hep-th},
  reportnumber  = {YITP-SB-18-27},
  doi           = {10.1088/1751-8121/aafd1b},
  journal       = {J. Phys. A},
  volume        = {52},
  number        = {7},
  pages         = {075401},
  year          = {2019}
}

@article{Fisher:1972zz,
  author  = {Fisher, Michael E. and Ma, Shang-keng and Nickel, B. G.},
  title   = {{Critical Exponents for Long-Range Interactions}},
  doi     = {10.1103/PhysRevLett.29.917},
  journal = {Phys. Rev. Lett.},
  volume  = {29},
  pages   = {917--920},
  year    = {1972}
}

@article{Paulos:2015jfa,
  author        = {Paulos, Miguel F. and Rychkov, Slava and van Rees, Balt C. and Zan, Bernardo},
  title         = {{Conformal Invariance in the Long-Range Ising Model}},
  eprint        = {1509.00008},
  archiveprefix = {arXiv},
  primaryclass  = {hep-th},
  reportnumber  = {CERN-PH-TH-2015-200},
  doi           = {10.1016/j.nuclphysb.2015.10.018},
  journal       = {Nucl. Phys. B},
  volume        = {902},
  pages         = {246--291},
  year          = {2016}
}

@article{DiPietro:2019hqe,
  author        = {Di Pietro, Lorenzo and Gaiotto, Davide and Lauria, Edoardo and Wu, Jingxiang},
  title         = {{3d Abelian Gauge Theories at the Boundary}},
  eprint        = {1902.09567},
  archiveprefix = {arXiv},
  primaryclass  = {hep-th},
  doi           = {10.1007/JHEP05(2019)091},
  journal       = {JHEP},
  volume        = {05},
  pages         = {091},
  year          = {2019}
}

@article{Billo:2016cpy,
  author        = {Bill{\`o}, Marco and Gon{\c{c}}alves, Vasco and Lauria, Edoardo and Meineri, Marco},
  title         = {{Defects in conformal field theory}},
  eprint        = {1601.02883},
  archiveprefix = {arXiv},
  primaryclass  = {hep-th},
  doi           = {10.1007/JHEP04(2016)091},
  journal       = {JHEP},
  volume        = {04},
  pages         = {091},
  year          = {2016}
}

@article{Kos:2015mba,
  author        = {Kos, Filip and Poland, David and Simmons-Duffin, David and Vichi, Alessandro},
  title         = {{Bootstrapping the O(N) Archipelago}},
  eprint        = {1504.07997},
  archiveprefix = {arXiv},
  primaryclass  = {hep-th},
  reportnumber  = {CERN-PH-TH-2015-097},
  doi           = {10.1007/JHEP11(2015)106},
  journal       = {JHEP},
  volume        = {11},
  pages         = {106},
  year          = {2015}
}

@article{Rong:2018okz,
  author        = {Rong, Junchen and Su, Ning},
  title         = {{Bootstrapping the minimal $ \mathcal{N} $ = 1 superconformal field theory in three dimensions}},
  eprint        = {1807.04434},
  archiveprefix = {arXiv},
  primaryclass  = {hep-th},
  doi           = {10.1007/JHEP06(2021)154},
  journal       = {JHEP},
  volume        = {06},
  pages         = {154},
  year          = {2021}
}

@article{Atanasov:2018kqw,
  author        = {Atanasov, Alexander and Hillman, Aaron and Poland, David},
  title         = {{Bootstrapping the Minimal 3D SCFT}},
  eprint        = {1807.05702},
  archiveprefix = {arXiv},
  primaryclass  = {hep-th},
  doi           = {10.1007/JHEP11(2018)140},
  journal       = {JHEP},
  volume        = {11},
  pages         = {140},
  year          = {2018}
}

@article{Heydeman:2020ijz,
  author        = {Heydeman, Matthew and Jepsen, Christian B. and Ji, Ziming and Yarom, Amos},
  title         = {{Renormalization and conformal invariance of non-local quantum electrodynamics}},
  eprint        = {2003.07895},
  archiveprefix = {arXiv},
  primaryclass  = {hep-th},
  reportnumber  = {PUPT-2613},
  doi           = {10.1007/JHEP08(2020)007},
  journal       = {JHEP},
  volume        = {08},
  pages         = {007},
  year          = {2020}
}

@article{Karateev:2025sjw,
  author        = {Karateev, Denis and Kravchuk, Petr and Manenti, Andrea and Reehorst, Marten and Vichi, Alessandro},
  title         = {{Bounds on Abelian Currents in 4d CFTs}},
  eprint        = {2512.20710},
  archiveprefix = {arXiv},
  primaryclass  = {hep-th},
  month         = {12},
  year          = {2025}
}

@article{Chester:2016wrc,
  author        = {Chester, Shai M. and Pufu, Silviu S.},
  title         = {{Towards bootstrapping QED$_{3}$}},
  eprint        = {1601.03476},
  archiveprefix = {arXiv},
  primaryclass  = {hep-th},
  reportnumber  = {PUPT-2494},
  doi           = {10.1007/JHEP08(2016)019},
  journal       = {JHEP},
  volume        = {08},
  pages         = {019},
  year          = {2016}
}

@article{Li:2018lyb,
  author        = {Li, Zhijin},
  title         = {{Bootstrapping conformal QED$_{3}$ and deconfined quantum critical point}},
  eprint        = {1812.09281},
  archiveprefix = {arXiv},
  primaryclass  = {hep-th},
  doi           = {10.1007/JHEP11(2022)005},
  journal       = {JHEP},
  volume        = {11},
  pages         = {005},
  year          = {2022}
}

@article{Landry:2019qug,
  author        = {Landry, Walter and Simmons-Duffin, David},
  title         = {{Scaling the semidefinite program solver SDPB}},
  eprint        = {1909.09745},
  archiveprefix = {arXiv},
  primaryclass  = {hep-th},
  reportnumber  = {CALT-TH 2019-038},
  month         = {9},
  year          = {2019}
}

\end{document}